\documentclass[10pt]{iopart}
\usepackage{color}
\usepackage{multirow}
\usepackage{mathtools}
\usepackage{diagbox}
\usepackage{iopams}
\usepackage{parskip}
\usepackage[utf8]{inputenc}
\usepackage{graphicx,subfigure}
\usepackage{cite,float,amsmath}
\usepackage{ulem}
\usepackage[english]{babel}
\usepackage{multicol}
\usepackage[colorlinks=true,allcolors=blue]{hyperref}

\begin{document}
\title[Validating neutral-beam current drive]{Validating neutral-beam current drive simulations in the TJ-II stellarator}

\author{S. Mulas$^1$, Á. Cappa$^1$, J. Martínez-Fernández$^1$,\\ D. López Bruna$^1$, J. L. Velasco$^1$, T. Estrada$^1$,\\ J. M. Gómez-Manchón$^1$, M. Liniers$^1$, K. J. McCarthy$^1$,\\
I. Pastor$^1$, F. Medina$^1$, E. Ascasíbar$^1$ and TJ--II Team}
\vspace{10pt}

\address{$^1$ Laboratorio Nacional de Fusión--CIEMAT, Madrid, Spain}

\ead{alvaro.cappa@ciemat.es}
\vspace{10pt}
\begin{indented}
\item[]January 2023
\end{indented}

\begin{abstract}
In this paper, we analyze the results of neutral-beam current drive (NBCD) experiments performed in the TJ-II stellarator with the aim of validating the theoretical predictions. Both parallel and anti-parallel injection with respect to the magnetic field were explored using co (NBI1) and counter (NBI2) beams at different injected beam power and plasma densities. The fast-ion current driven by both beams was simulated with the Monte Carlo code ASCOT and the electron response to the fast-ion current was calculated analytically using a model valid for an arbitrary magnetic configuration and a low collisionality plasma. Despite the uncertainties associated to the determination of experimental inputs, the model reproduces with rather good agreement the toroidal current measured in NBI2 plasmas. However, the current driven by NBI1 is less than half the predicted one. Possible reasons for this discrepancy are discussed. Among the probable causes, yet to be studied, the most likely is the increased presence of lithium in the plasma when NBI1 is injected, this being the result of its irregular deposition during wall conditioning.       
\end{abstract}

\maketitle
\ioptwocol
 
\section{Introduction}

Progress in understanding fast ion slowing down and confinement in fusion devices necessarily involves the validation of the available numerical tools against the experimental observations. Fast ions produced by neutral beam injection (NBI) systems are key to this goal. Besides the results related to neutral beam power deposition and heating performance, the current induced by the injection of fast ions is an experimentally measurable quantity that may be derived from slowing down simulations and analytic electron response model calculations \cite{Lin-Liu_1997, Hu_2012, Nakajima_1990, Mulas_2022}, making it a perfect candidate for model validation.

Unlike the body of work done in tokamaks \cite{Oikawa_2000, Politzer_2005, Park_2009, Murakami_2009, Suzuki_2011}, there is not much literature dedicated to the experimental study of neutral beam current drive (NBCD) in stellarators \cite{Nagaoka_2006, Lazerson_2021}, and even less to theory validation. Moreover, having a validated tool to estimate NBCD in non-axisymmetric configurations is also desirable when interpreting the results of experiments studying Alfvén Eigenmodes. In fact, validating NBI-driven Alfvén Eigenmodes (AEs) simulations requires, among other inputs, knowing the rotational transform profile of the plasma equilibrium, which is strongly affected by non-inductive plasma currents such as NBCD. In particular, in the case of the TJ-II stellarator, and since many of the experiments have been carried out with non-balanced neutral beam injection and no measurements of the rotational transform profile are available (or they present a large error in those few cases where it has been measured using motional Stark effect (MSE) diagnostic \cite{McCarthy_2015}), a theoretical estimate of the current driven by the injection of the neutral beam is needed to reconstruct the rotational transform profile \cite{Cappa_2021}. This has been the initial drive for the NBCD validation studies in TJ-II presented in this paper. The beam driven current is the combination of two different contributions: the current of the fast ions and the response of the plasma electrons that shield this current. In a previous paper \cite{Mulas_2022}, the current associated with the slowing-down distribution of NBI driven fast ions in TJ-II was obtained using the Monte Carlo orbit-following code ASCOT \cite{Hirvijoki_2014} and the return or shielding current of the electrons was determined using the derivation presented in appendix A of reference \cite{Mulas_2022}. The simulations presented in \cite{Mulas_2022}, performed for high density plasmas, did not include charge exchange (CX) processes. In devices with higher temperatures and higher neutral beam injection energies, CX losses are lower due to the strong decrease of the CX cross section as energy increases, and therefore usually neglected. However, in the case of TJ-II NBI plasmas, in particular if they are interesting from the point of view of AEs studies (relatively low plasma density), the density of neutrals atoms in the plasma makes the CX losses relevant and they need to be included if a reliable estimate of the fast ion slowing down distribution is desired \cite{Ollus2022}. In low density conditions, former guiding center calculations done with FAFNER \cite{Lister_1985} showed that CX losses in TJ-II plasmas can reach up to 30\% of the port through power, which represents almost 70\% of the available power due to the high level of shine through \cite{GuaspIR_1995}. In this paper, we will use a version of the ASCOT code that includes CX processes to calculate the slowing-down distribution function, which is the necessary input to estimate the amount of current driven by the beams.

In general, the measured toroidal plasma current does not reach a stable value during the plasma shot because the L/R time ($\tau_{LR}$) is often of the order of the typical TJ-II NBI pulse length (100 ms) and, therefore, to validate the numerical results, we need an estimate of the stable asymptotic current that would be achieved with longer pulses. Moreover, since the total current is actually a combination of the neutral beam driven current and the bootstrap current, an estimate of the latter, provided by the DKES code \cite{Hirshman_1986, Velasco_2011}, is also needed in order to isolate the NBCD contribution.

We must face an additional complication in the case of TJ-II stellarator; because of the proximity of the main field conductors to the plasma, even a small level of ripple in the currents through the coils induce unwanted oscillations in the plasma current ($\approx 0.1-0.3$ kA). Removing these oscillations considerably improves the time evolution of the plasma current and allows to extrapolate its asymptotic level with a higher degree of accuracy. \ref{ap:a} presents the method used to determine the underlying values of the plasma current due only to the non-inductive current sources. In addition, infrared measurements of the NBI shine-through power are available and help us to confirm the injection geometry and the estimates of available NBI power in the plasma given by the simulation.

The paper is organized as follows. In section \ref{sec:exp}, after a brief description of the NBI system, we present the experimental results obtained with each of the two injectors, including the radial profiles of the main measurable quantities needed for the NBI simulations and the measured toroidal currents. Section \ref{sec:sim} is divided in different subsections. Subsection \ref{sec:shine} is devoted to the comparison of the measured and calculated beam power shine-through; the results of the slowing-down simulations are presented in subsection \ref{sec:ascot} and finally, the neutral beam current drive calculation and the comparison with the experimental values is performed in subsection \ref{sec:nbcd}.    

\section{Experimental results}
\label{sec:exp}

The NBI system of the TJ-II stellarator (on-axis field $B_0=0.95$ T, $a\leq 0.22$ m and $R=1.5$ m) consists of two tangential hydrogen beams (co and counter) injecting 700 kW of maximum power each with a maximum energy of 34 keV \cite{Liniers_2017}. 
\begin{figure}[h]
\begin{center}
\includegraphics[width=0.45\textwidth]{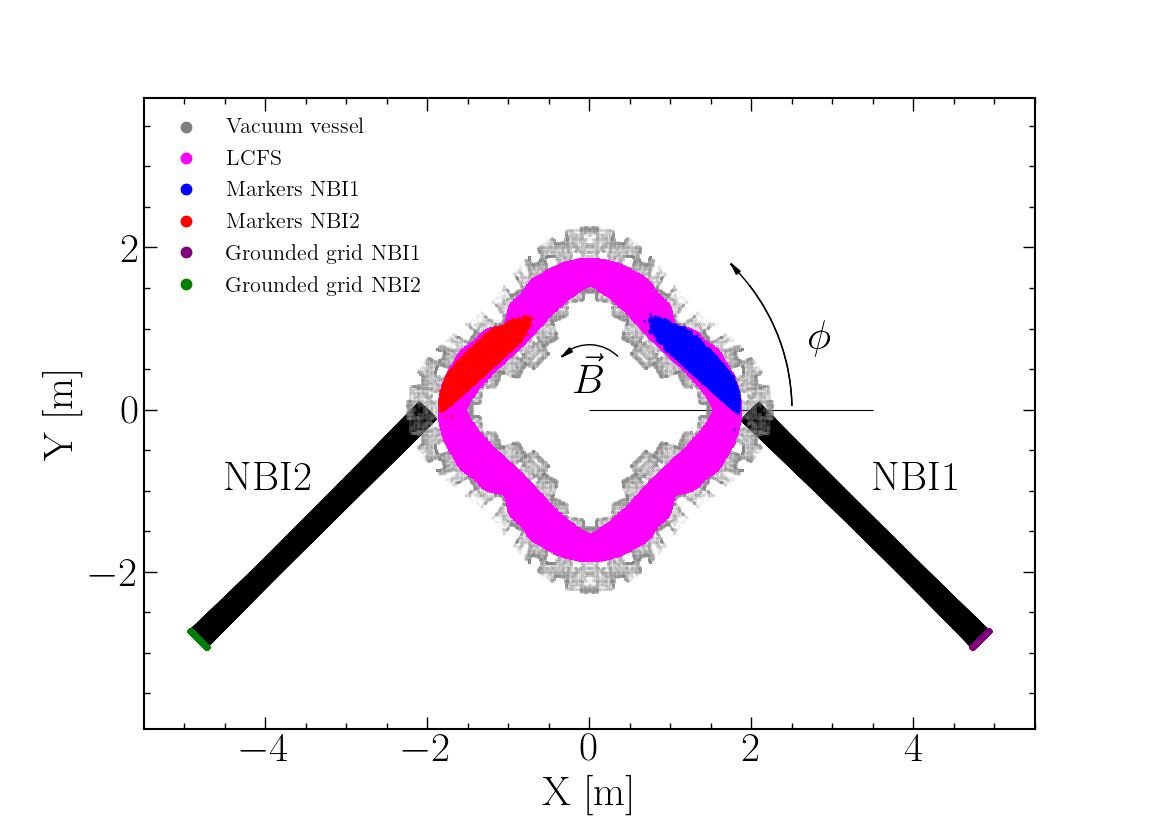}
\end{center}
\caption{NBI system injection geometry. The cloud of initial markers representing birth ions are shown for both injectors.}
\label{fig:NBIsys}
\end{figure}
Figure \ref{fig:NBIsys} illustrates the injection direction of each neutral beam, the main direction of magnetic field, and the spatial distribution of birth ions created by both injectors, which has been calculated with the Beamlet-Based Neutral-Beam Injection (BBNBI) module of the ASCOT code \cite{Asunta_2015} (see section \ref{sec:sim} for details).

To proceed, a set of four selected hydrogen plasmas with neutral beam injection have been chosen from the TJ-II database. One is an old shot with good density control during the NBI phase (\#24000) and the other three (\#53577, \#53605 and \#54097) have been taken from a larger set of shots specifically designed to investigate NBCD. They all exhibit approximately stationary electron line density and temperature during the NBI phase. This guarantees that the source of driven current is roughly constant and that, once the externally driven inductive components are removed (see \ref{ap:a}), the time evolution of plasma current measured experimentally is only originated by the plasma self-inductive response to the internal current sources (NBCD and bootstrap). 

\begin{figure}[h]
\begin{center}
\includegraphics[width=0.45\textwidth]{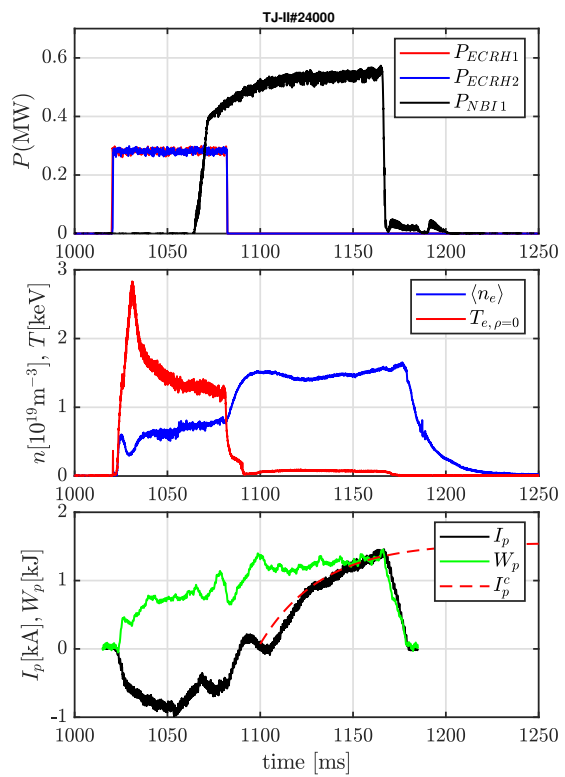}
\end{center}
\caption{Time traces of heating power (top panel), line density and central ECE temperature (central panel), plasma energy and toroidal current (bottom panel). The toroidal current with ($I_p$) and without ($I_p^c$) externally driven inductive contributions is shown. Here, only NBI heating is used (no ECRH). Note that during the high density NBI phase, the cutoff density of the X2 mode is exceeded in the plasma core (for $t>1090$ ms) and central ECE channels stop measuring.}
\label{fig:NBI_plasma_traces}
\end{figure}

\begin{figure}[h]
\begin{center}
\includegraphics[width=0.45\textwidth]{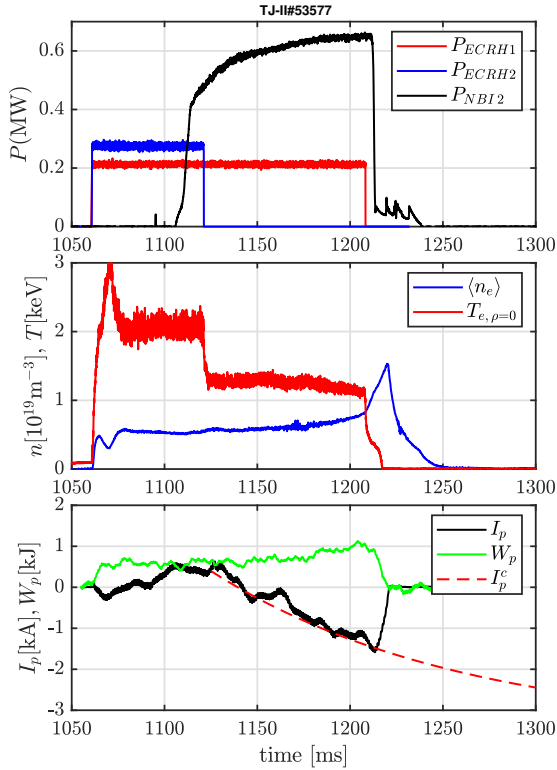}
\end{center}
\caption{Same as figure \ref{fig:NBI_plasma_traces} for a case in which ECRH is applied during the whole shot to control the plasma density. Here, NBI2 is used.}
\label{fig:NBIECH_plasma_traces}
\end{figure}

Figures \ref{fig:NBI_plasma_traces} and \ref{fig:NBIECH_plasma_traces} show the time evolution of the plasma parameters and the heating power for two of the four representative shots. Stable density plasmas heated only by NBI are difficult to achieve in TJ-II and the number of suitable discharges for NBCD studies is limited. Only right after an optimum lithium wall conditioning, the line density can be kept constant enough so as to ensure that the time evolution of the plasma current has only an electrodynamic origin characterized by $\tau_{LR}$. 

In all cases, second harmonic ECRH is used to start-up and build the NBI target plasma and, in situations with difficult density control, ECRH is used to maintain a constant density during the NBI phase. The NBI operation parameters, beam voltage and port-through power, as well as the central plasma density, for each of the shots studied, have been collected in table \ref{tab:nbiparam}.

\begin{table}[h]
    \small
	\centering
	\begin{tabular}{|c|c|c|c|c|}
		\hline
		\multicolumn{5}{|c|}{\textbf{NBI parameters and central density}}\\
		\hline
		\textbf{Inj.} & \textbf{\#shot} &\textbf{$V_b$ [kV]}& \textbf{P [kW]}& \textbf{$n_0$[$10^{19}$m$^{-3}$]}\\ 
		\hline\hline
		 \multirow{2}{*}{NBI2} & 53577* & 29.5 & 477 & 0.9 \\
                               & 53605* & 27.1 & 324 & 0.9\\
		\hline
		\multirow{2}{*}{NBI1} & 54097* & 27.8 & 286 & 0.9\\
		                      & 24000 & 32   & 430 & 2.4 \\ 
		\hline
	\end{tabular}
	\caption{NBI parameters used in the four cases analyzed. Shot numbers marked with an asterisk (*) indicate the use of ECRH during the NBI phase.}
	\label{tab:nbiparam}
\end{table}

For the case represented in figure \ref{fig:NBI_plasma_traces}, provided that no EC current is being driven, the only source of current in the plasma during the ECRH phase is the bootstrap current, which is driven by the plasma gradients.

After switching on the NBI, the ECRH is switched off and the plasma current, now driven by NBI,  starts evolving towards an asymptotic steady-state value. In this case, the co-NBI injector is used (NBI1) and the current during the NBI phase is positive. Figure \ref{fig:NBIECH_plasma_traces} illustrates a case with the counter-injector (NBI2) in which perpendicular ECRH is successfully used to control the plasma density. This, on the other hand, limits the range of achievable densities. In this case, the current in the NBI phase is negative. As mentioned in the introduction, the time behaviour of the plasma current always exhibits slow oscillations of the order of 0.1-0.3 kA due to the time dependent ripple of the currents in the main field coils. Following the optimization procedure described in \ref{ap:a}, we can calculate the time evolution of the plasma current, once the externally driven inductive contributions have been extracted. The result is shown with the dashed red line ($I_p^c$) in figures \ref{fig:NBI_plasma_traces} and \ref{fig:NBIECH_plasma_traces}. The asymptotic value of this curve, $I_{ni}$, which is given by the optimization algorithm, is the one that we must compare with the plasma current provided by the simulations. The L/R time ($\tau_{LR}$) is also a result of the optimization procedure.    

Core radial profiles of electron plasma density and temperature are measured by Thomson scattering while edge profiles are obtained from profile reflectometry \cite{Estrada_2001} and helium beam \cite{Hidalgo_2004} measurements. 

\begin{figure}[h]
\begin{center}
\includegraphics[width=0.38\textwidth]{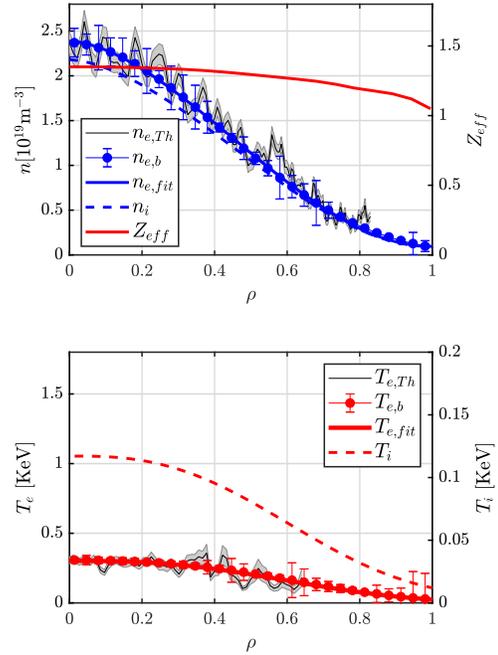}
\end{center}
\caption{Thermal plasma profiles measured at $t=1120$ ms (during the NBI phase) for shot \#24000.}
\label{fig:NBI_profiles}
\end{figure}

Bayesian analysis \cite{VanMilligen_2011}, that considers also the measured value of interferometric line integrated density, is used to reconstruct the full radial profiles. Ion density is deduced from the effective charge ($Z_{eff}$) profiles calculated from radiation measurements. Typical values of $Z_{eff}$ below 2 are measured in NBI plasmas. These values are estimated from the emission of SXR recorded by four detectors with four different beryllium filters. Using the electron density and temperature profiles, a simulation of the detected emission is done by using the radiation code IONEQ. The uncertainty of these calculations is difficult to ascertain, being 20\% a rough estimate of its upper limit.

\begin{figure}[h]
\begin{center}
\includegraphics[width=0.38\textwidth]{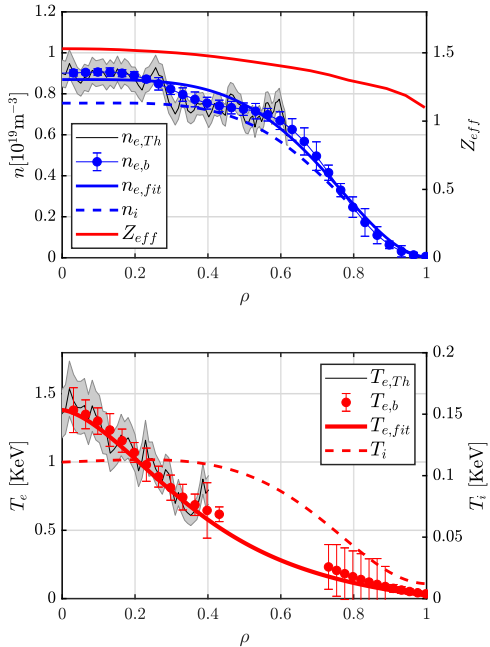}
\end{center}
\caption{Thermal plasma profiles measured at $t=1120$ ms for shot \#53577.}
\label{fig:NBIECH_profiles}
\end{figure}

The central ion temperature is measured with the neutral particle analyzer (NPA) and its radial profile is inferred using an approximation developed in \cite{Fontdecaba_2010}. Figure \ref{fig:NBI_profiles} shows the radial profiles of these quantities measured in the stable phase of the NBI high density plasma shown in figure \ref{fig:NBI_plasma_traces} while figure \ref{fig:NBIECH_profiles} shows the same data for one of the the low density NBI+ECRH shots ($\rho=\sqrt{s}$ is the radial coordinate, being $s$ the normalized toroidal flux). In both figures, the shaded grey region indicates the Thomson raw data errors bars while the blue (density) and red (temperature) dots, as well as their associated error bars, are the output of the Bayesian analysis. For simulation purposes, and also to cope with the lack of Thomson data in low density regions due to the signal being dominated by parasitic laser light, smoothed fitted data (solid red and blue lines) are used. 

As anticipated in the introduction, including neutral profiles to simulate charge exchange reactions is essential to achieve a reliable estimate of the fast ion slowing-down distribution. The probability of neutralization of fast ions inside the plasma depends on the density of neutrals found along their paths. In order to simplify the problem, we have obtained average radial profiles of atoms and molecules out of the calculated 3D distributions in the volume given by the magnetic configuration. The 3D calculations have been obtained with the Monte Carlo code EIRENE adapted to the geometry and characteristics of the TJ-II stellarator. Since one of the main parameters for these calculations is the particle confinement time, we have used previous experience on similar plasmas to estimate the neutrals considering a wall recycling factor around $0.7$. 

\begin{figure}[h]
\begin{center}
\includegraphics[width=0.42\textwidth]{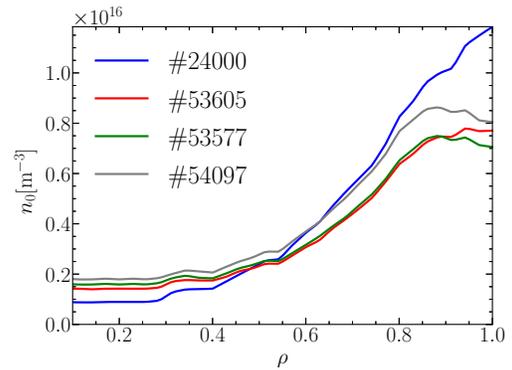}
\end{center}
\caption{Radial profiles of neutral atoms.} 
\label{fig:neutral_profiles}
\end{figure}

Self-consistent transport calculations are done for the electron density using the ASTRA suite \cite{Lopez-Bruna_2013, Lopez-Bruna_2018} coupled to peripheral codes \cite{Lopez-Bruna_2010} for the slow (EIRENE) and fast (FAFNER \cite{Lister_1985}) neutrals, both of which intervene in the electron transport problem. The transport coefficients are adjusted so as to approximately reproduce the electron density profiles in steady state with all the particle sources. This provides electron confinement times compatible with those expected for the given operation conditions ($\sim 5$ ms) and the corresponding neutrals distributions.

\begin{figure}[h]
\begin{center}
\includegraphics[width=0.42\textwidth]{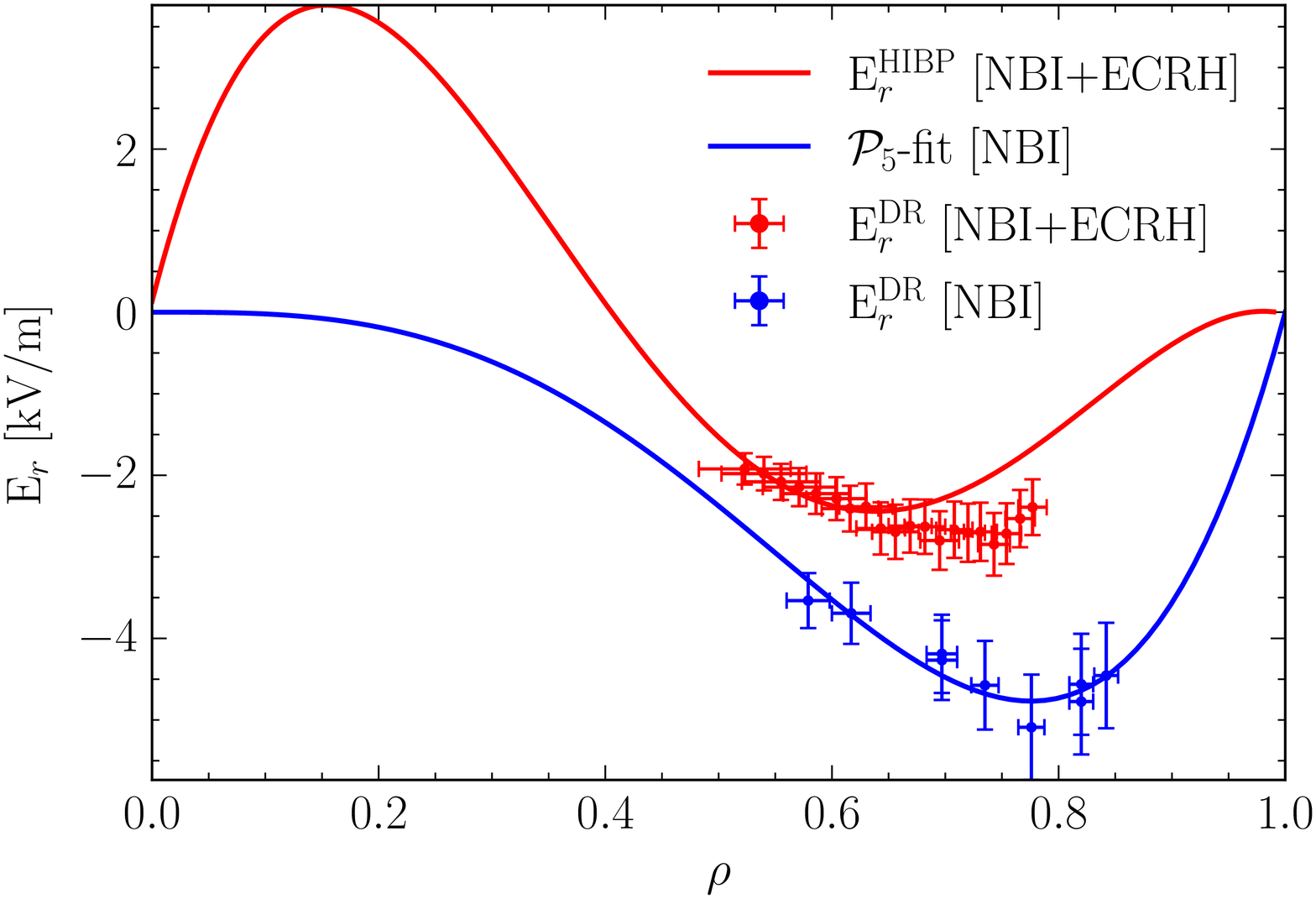}
\end{center}
\caption{Radial electric field for the low (red) and medium density plasmas (blue). Doppler reflectometry measurements are shown with solid dots. HIBP data (represented by a fit to the experimental measurements) was only available for the low density NBI+ECRH case. The fit for the NBI case is done knowing that at $\langle n_e\rangle \approx 1.5\times10^{19}$ m$^{-3}$ the field is negative for all values of $\rho$ since the plasma is in ion root (see text).}
\label{fig:er_profECRH}
\end{figure}

Finally, although its impact on fast ion slowing-down is minor \cite{Mulas_2022}, the electric field has also been included in the slowing-down simulations. In TJ-II, the plasma potential is measured by radially scanning the heavy ion beam probes \cite{Melnikov_2022}. From this measurement, a radial profile of the electric field across the whole plasma column can be estimated. Moreover, Doppler reflectometry also provides a measurement of the radial electric field in the outer region of the plasma \cite{Happel_2009}. For the low density NBI+ECRH plasmas studied in this paper (obtained in shots \#53577, \#53605 and \#54097), both HIBP and Doppler reflectometry data were available and have been used to reconstruct the radial profile of the electric field. For the medium density case (shot \#24000) only Doppler reflectometry data were available. Electric field profiles are shown in figure \ref{fig:er_profECRH}. The measured profiles are consistent with the findings described in previous studies for plasmas in electron root (positive field in the plasma core) and ion root (negative field in all the plasma column) \cite{Estrada_2009, Arevalo_2014, Melnikov_2022}.

\section{NBI simulations and comparison with experimental results}
\label{sec:sim}

The injection of fast NBI hydrogen neutrals into the plasma and the distribution of birth ions has been simulated with BBNBI. The maximum energy of the newly born ions and the port-through (PT) power in each simulation correspond to those of table~\ref{tab:nbiparam}. The fractions of injected particles at different energies, that is, $E_{max}=eV_{b}$, $E_{max}/2$ and $E_{max}/3$, have been obtained from spectroscopic measurements \cite{McCarthy2010} in the NBI ducts, being 48\%, 24\% and 28\% for NBI1 and 38\%, 30\% and 32\% for NBI2 respectively. The standard TJ-II configuration has been used in all the experiments and its corresponding VMEC vacuum equilibrium used in the simulations. The magnetic field outside the last closed flux surface (LCFS), that is needed by ASCOT to take into account properly the particle trajectories, was extended until the first wall with the EXTENDER code \cite{Drevlak_2005}. 

\subsection{Shine through}
\label{sec:shine}

From the results of the simulations with BBNBI, we can estimate the power loads to the wall due to shine through (ST) and compare with the measurements taken by an infrared camera (IRC) recently commissioned in TJ-II for this purpose \cite{Liniers_2019}. In principle, the experimental temperature rise in the vacuum vessel at the beam impact location (beam stop) due to a continuous injection is given by
\begin{equation}
T(t)=T_0+\frac{q}{C(T)}\sqrt{t},
\label{eq:eq1}
\end{equation}
where $T_0$ is the initial temperature of the target, $q$ is the heat flux, and $C(T)$ is a temperature-dependent coefficient specific of the material of the target \cite{Cornelissen_2022}. The term $C(T)$ can not be estimated due to the uncertainty in the composition of the beam stop because of the recurrent treatment of the first wall with boron and lithium to avoid sputtering from the walls and facilitate density control. 
\begin{figure}[t]
	\begin{center}
		\includegraphics[width=0.5\textwidth]{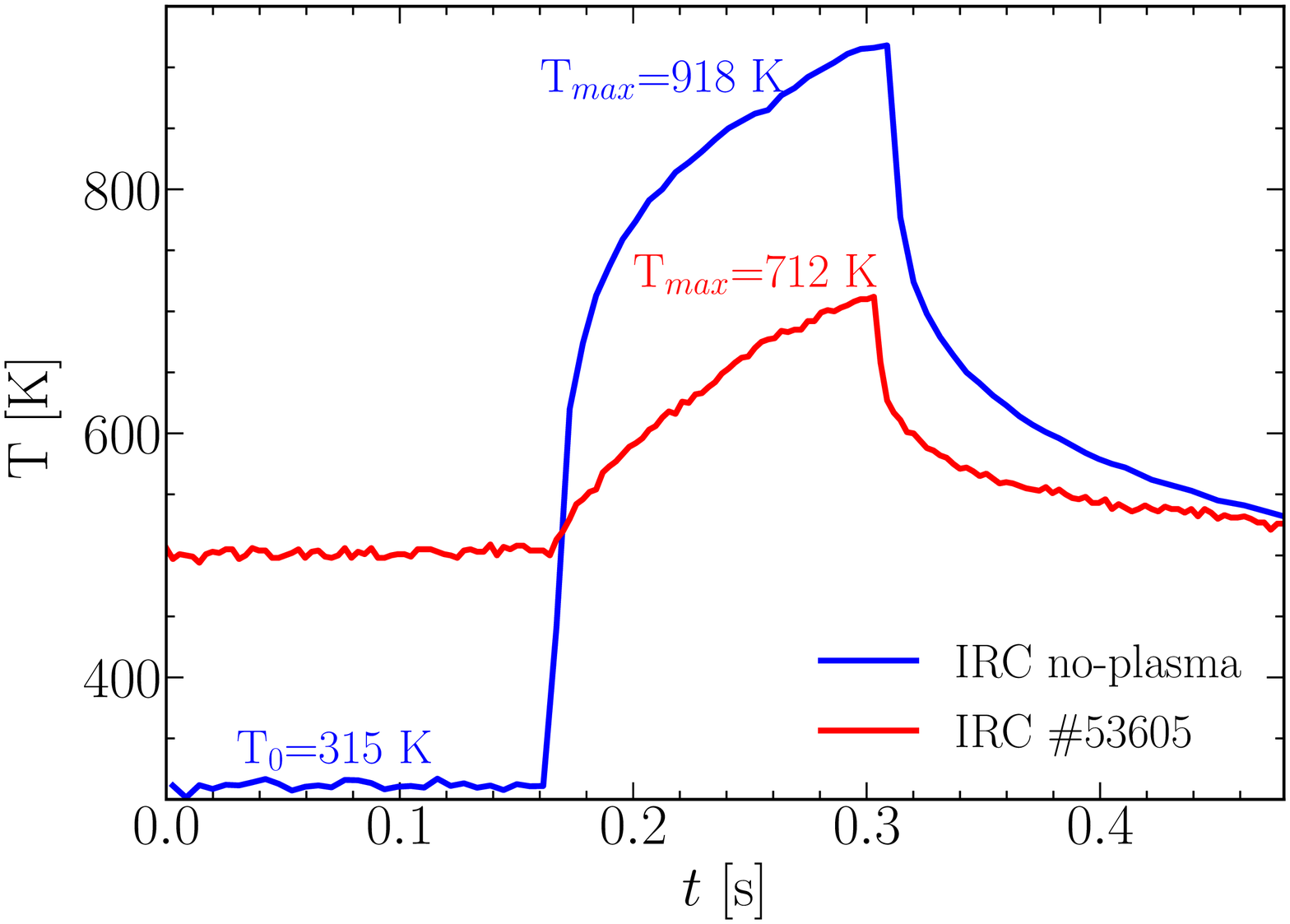}
	\end{center}
	\caption{Beam-dump temperature from IRC signal during shot \#53605 (red line) and a calibration shot without plasma (blue line). A high-temperature filter only allowed measurements above 500 K in shot \#53605. The initial temperature before the NBI pulse was 315 K.}
	\label{fig:IRCtraces}
\end{figure}
To sort out this uncertainty, the expression used to obtain the ST-power fraction out of the injected power is
\begin{equation}
f_{ST}\coloneqq\frac{q^{shot}}{q^{vac}}=\frac{T^{shot}_{max}-T^{shot}(t_0)}{T^{vac}_{max}-T^{vac}(t_0)}.
\label{eq:Fst}
\end{equation}
Here, $t_0$ is the NBI switch-on time, the superscripts refer to the values measured either during the actual shot (shot) or during a calibration shot without plasma (vac), see figure \ref{fig:IRCtraces}. Even if equation \eqref{eq:eq1} were not fulfilled because of the unknown thermal properties of the graphite target, heavily covered with lithium compounds, beam power scans without plasma have proved that the increments in temperature depend linearly on the injected power and therefore equation \eqref{eq:Fst} still holds.  

For the cases with available data, the values obtained both with BBNBI and experimentally (equation \ref{eq:Fst}) are in close agreement and prove to be quite high, specially in the low-density NBI+ECRH plasmas (see Table \ref{tab:shinedpower}). 
\begin{table}[t]
	\centering
	\begin{tabular}{|c|c|c|c|} 
		\cline{3-4}
		\multicolumn{2}{c|}{} & \multicolumn{2}{c|}{\textbf{Shine-through [\%]}}\\
		\hline
		\textbf{Injector} & \textbf{\#Shot} & \textbf{BBNBI} & \textbf{IRC}\\ 
		\hline\hline
		\multirow{2}{*}{NBI2} & 53577 & 66 & 67\\
		& 53605 & 66 & 66\\
		\hline
		\multirow{2}{*}{NBI1} & 54097 & 72 & -\\
		& 24000 & 55 & 49\\ 
		\hline
	\end{tabular}
	\caption{Shine-through as calculated by BBNBI and measured with infrared camera (IRC calibration shot was not available for the NBI power used in shot \#54097).}
	\label{tab:shinedpower}
\end{table}
Port-through power, corrected with the  shine-through values, determines the available power in each case. These are shown in table \ref{tab:powerdis}. As for the shine-through value  taken for shot \#24000 (no IRC data were available at that time), it has been actually measured in a plasma with same line density and NBI1 heating power.

\begin{figure}[h]
	\begin{center}
		\subfigure[]{\includegraphics[width=0.225\textwidth,height=0.1875\textwidth]{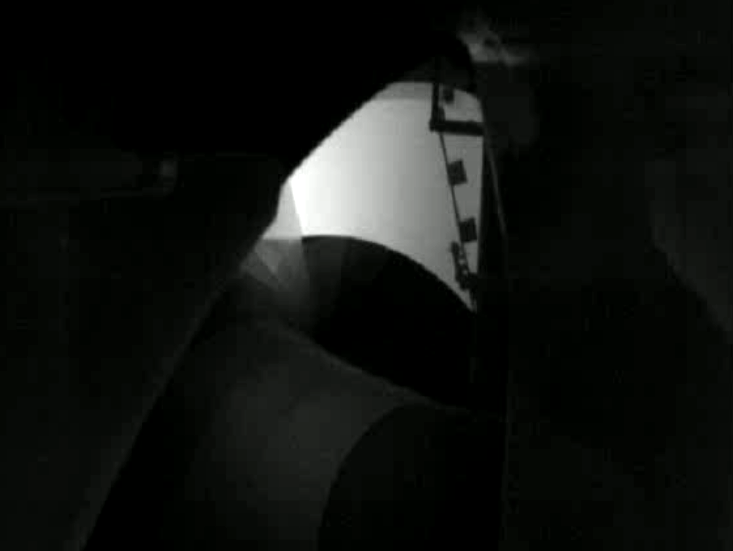}\label{fig:IRC0}}\hspace{0.0cm}
		\subfigure[]{\includegraphics[width=0.225\textwidth,height=0.1875\textwidth]{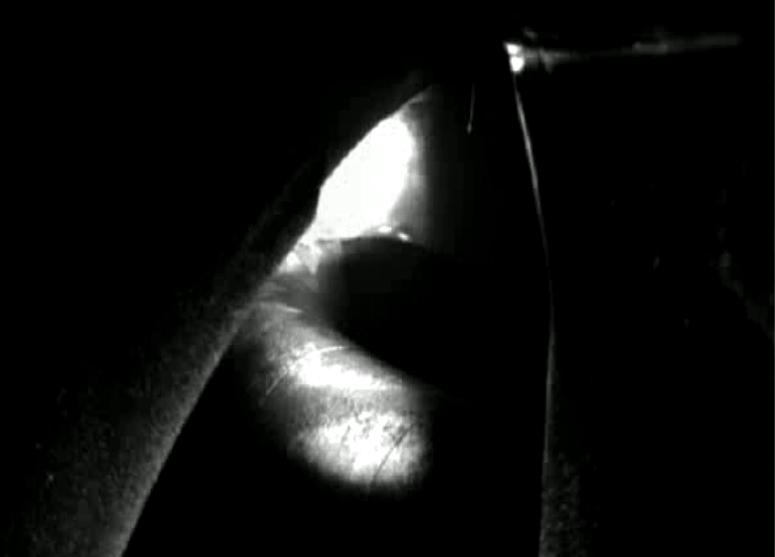}\label{fig:IRC1}}\\
		\subfigure[]{\includegraphics[width=0.225\textwidth,height=0.1875\textwidth]{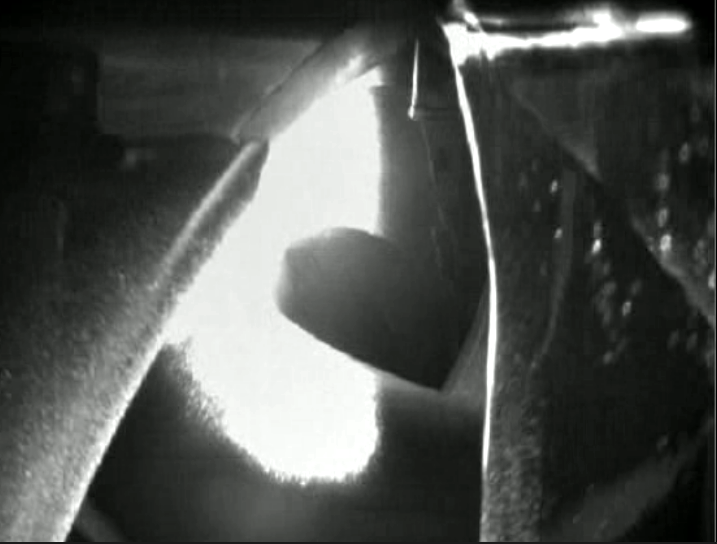}\label{fig:IRC2}}\hspace{0.0cm}
		\subfigure[]{\includegraphics[width=0.225\textwidth,height=0.1875\textwidth]{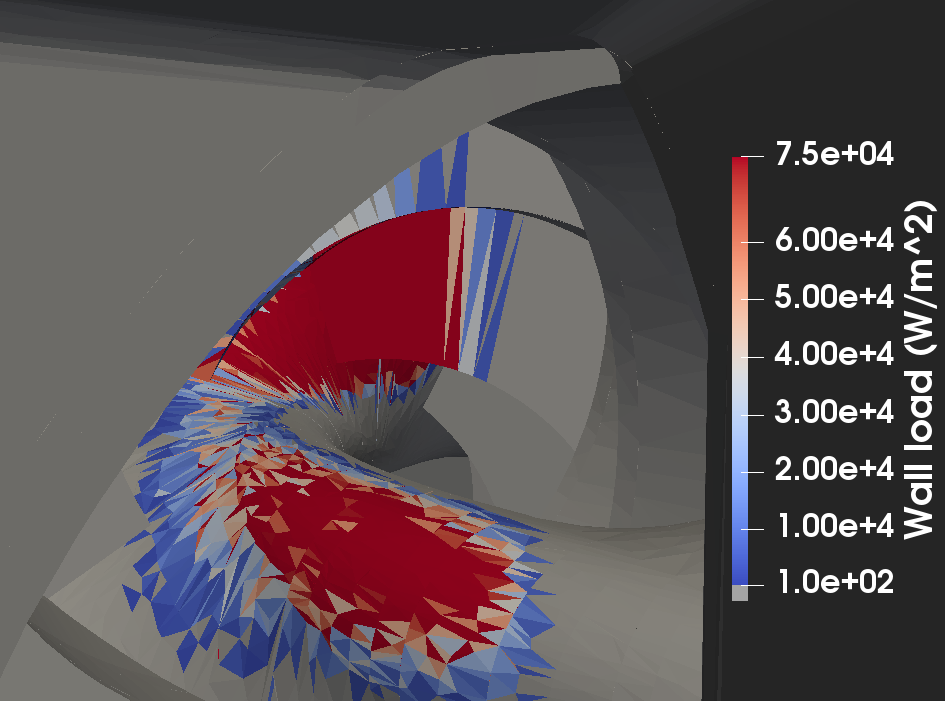}\label{fig:BBNBI1}}
	\end{center}
	\caption{IRC images taken at different times during shot \#54097 (figures a, b and c) and ST power load on the vessel as calculated by BBNBI (d). The colorbar is not covering the full range of values to show better the qualitative agreement of the simulation with the IRC data. The maximum heat load obtained in the simulation is $0.5$ MW/m$^2$}.
	\label{fig:STcomparison}
\end{figure}

In figure \ref{fig:STcomparison}, IRC images at different times during the NBI1 pulse in shot \#54097 are presented together with the ST-power loads as calculated by BBNBI. There is a good agreement, not only between experimental and calculated ST values, but also between measured and calculated power distribution loads.   

\subsection{ASCOT simulations}
\label{sec:ascot}

The dynamics of fast ions taken into account in the simulations has three different contributions: the Hamiltonian motion due to the electromagnetic field, the slowing down and pitch-angle scattering produced by collisions with the bulk plasma, and the stochastic CX reactions. The relative importance of these mechanisms can produce quite different steady-state fast-ion distribution functions. At fixed available power, a more collisional plasma (i.e lower temperatures and higher densities) would reduce the fast-ion density because of the earlier thermalization and a more intense pitch-angle scattering that enhances the transition of particles from passing to trapped regions of phase space where they are quickly lost. Likewise, low fast-ion energies and high neutral densities would reduce the population of fast ions due to a strong diffusion of neutralization/reionization cycles by CX reactions. 

\begin{figure}[h]
	\begin{center}
		\subfigure[]{\includegraphics[width=0.4\textwidth]{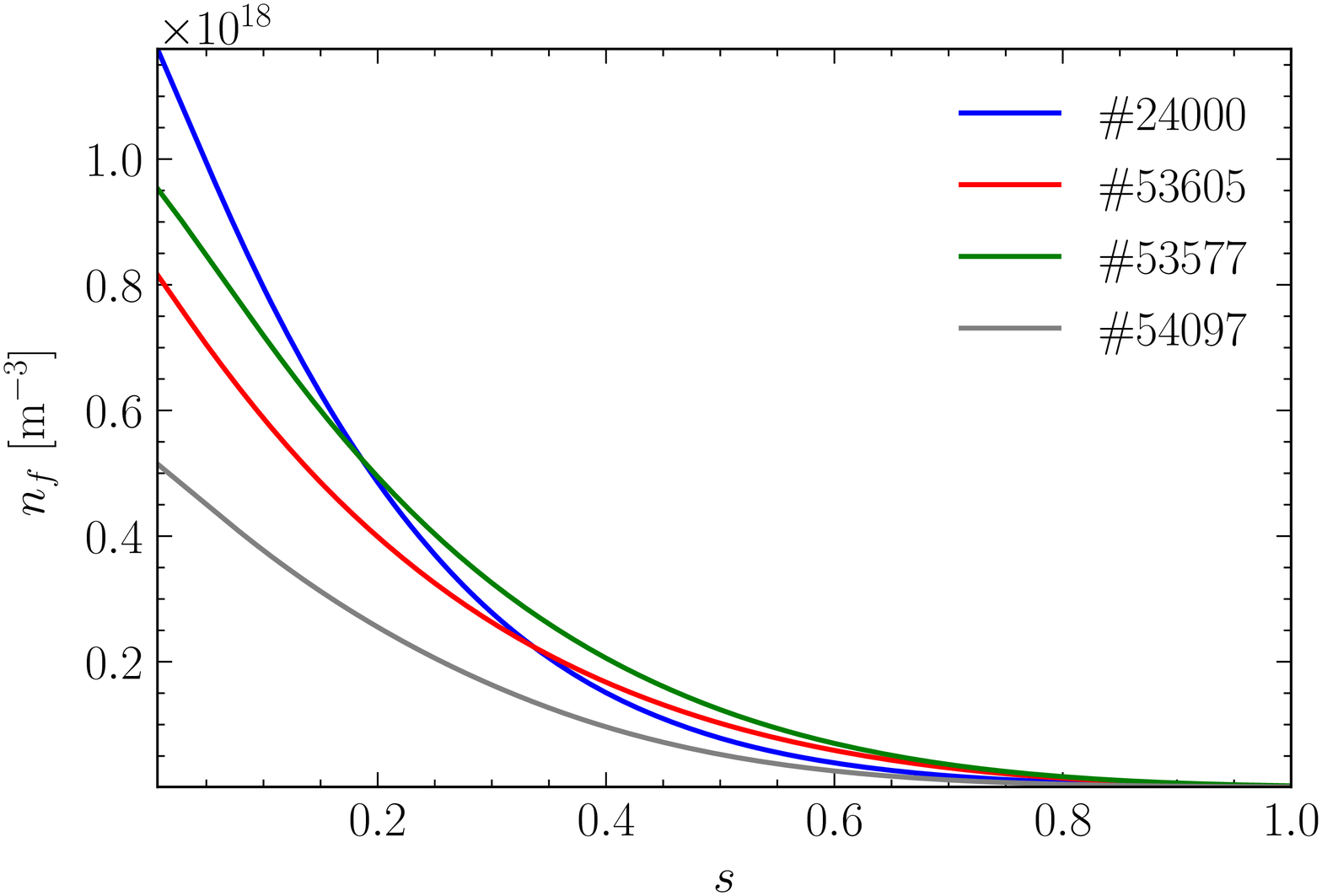}\label{fig:FIdens}}
    	\subfigure[]{\includegraphics[width=0.4\textwidth]{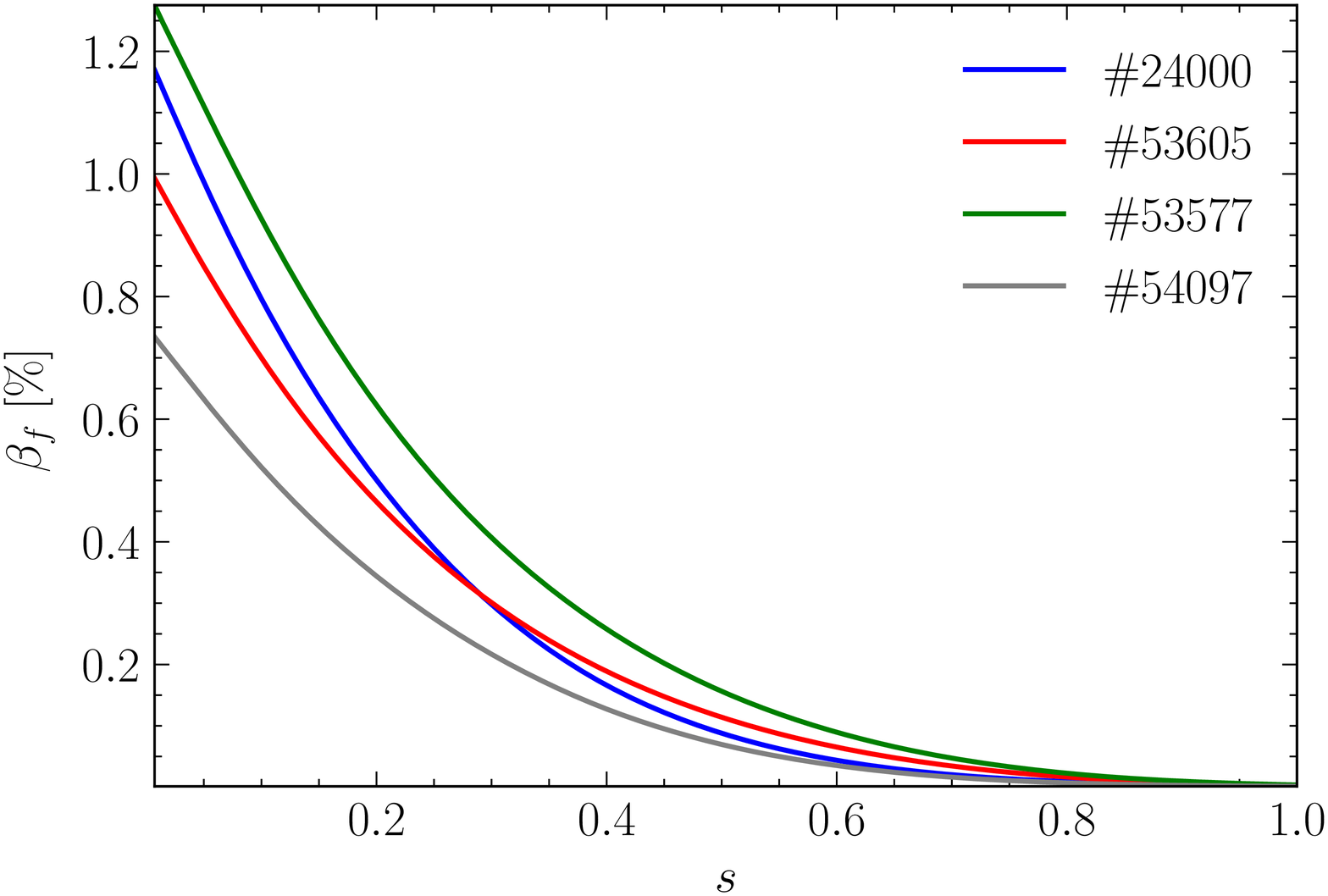}\label{fig:FIbetas}}
    	\end{center}
	\caption{Fast-ion density \protect\subref{fig:FIdens} and fast-ion $\beta$ profiles \protect\subref{fig:FIbetas}, calculated by ASCOT.}
	\label{fig:FIdensbeta}
\end{figure}

\begin{figure}[h]
	\begin{center}
		\includegraphics[width=0.4\textwidth]{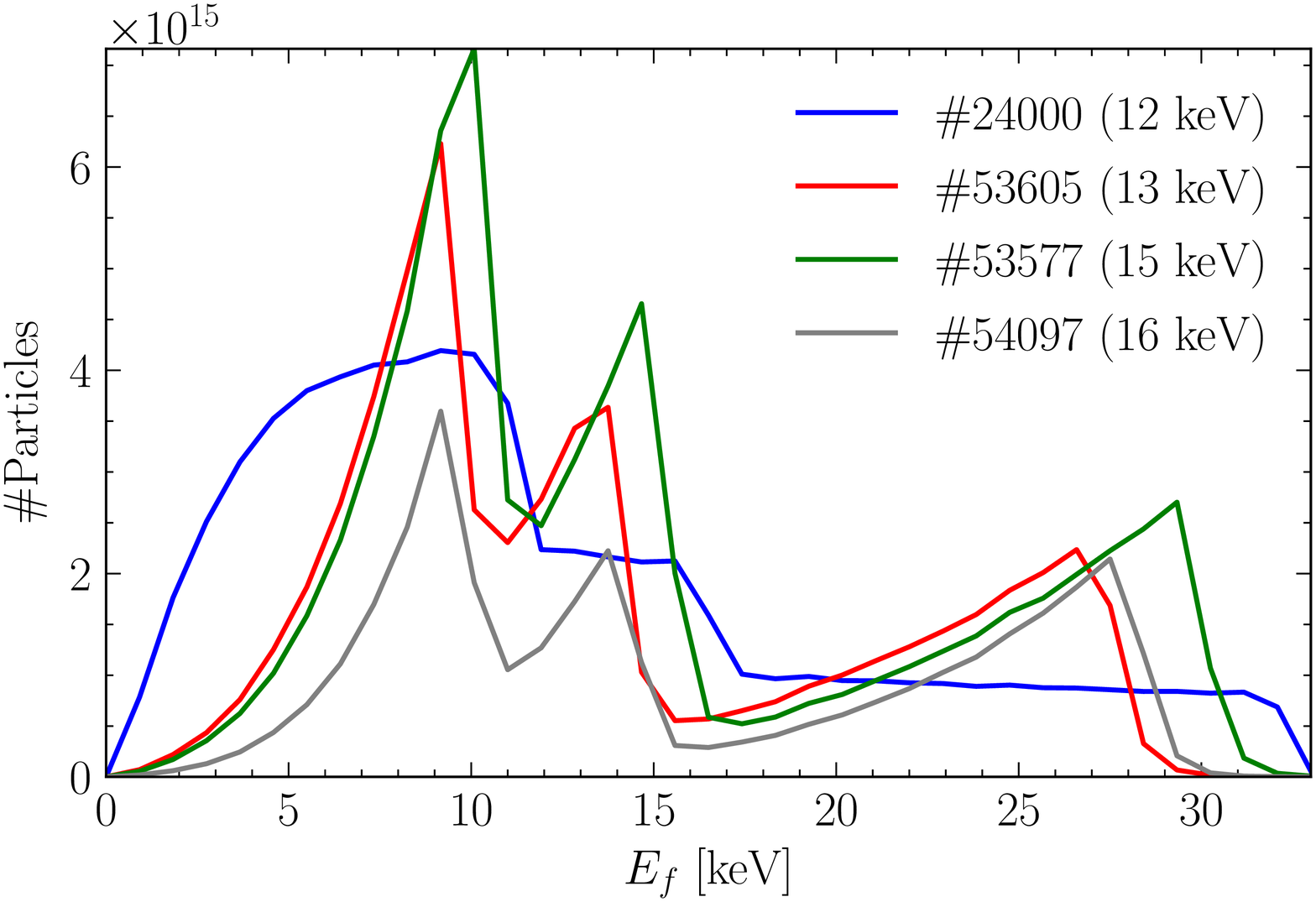}
    	\end{center}
	\caption{Fast-ion energy distributions. The numbers in parenthesis are the average energies of the distributions.}
	\label{fig:energydists}
\end{figure}

The computed fast-ion density profiles, $n_f(\rho)$, shown figure~\ref{fig:FIdens}, are approximately one order of magnitude lower than their corresponding plasma densities. The fast-ion densities are ordered accordingly to the available powers (see table \ref{tab:powerdis}) being shot \#24000 the one with the highest fast-ion density despite being the most collisional plasma, as a result of the more efficient ionization of NBI neutrals due to the higher plasma density. On the contrary, this shot presents values of fast-ion pressure ($\beta_{f}=p_{f}/p_{mag}$, where $p_{f}=n_{f}E_{f}$ and $p_{mag}=\langle B^2\rangle/2\mu_0$ are the fast-ion pressure and the magnetic pressure respectively) similar to those of the other shots. The average energy of the fast-ion distribution of shot \#24000 is lower than in the other shots (see figure \ref{fig:energydists}), which compensates for the higher fast-ion density and results in $\beta_{f}$ values closer (and even lower) to the  ones obtained in the other shots. This lower average energy is caused by the presence of a larger amount of slowed-down particles in the distribution (e.g particles with energy below 10 keV), which are missing in the other shots due to the dominance of CX reactions. In fact, almost only newly born fast ions populate the distribution functions of the low density shots (notice the decay of the number of particles as energy decreases, starting at each of the injection energies of the NBI). The qualitative difference between the shapes of the energy distribution for the low-density plasmas and the medium-density one reflects the fact that, while the time scale of the CX reactions is comparable for all the shots (the time scale only depends on the CX cross section, fast-ion energy and neutral density, which are similar in all cases), the slowing-down time is much higher for the low-density cases due to their lower collisionality. This causes that fast ions hardly have enough time to thermalize before being lost by CX reactions. 

\begin{table}[h]
	\centering
	\begin{tabular}{|c|c|c|c|c|c|} 
		\cline{3-6}
		\multicolumn{2}{c|}{} & \multicolumn{4}{c|}{\textbf{Power balance [kW]}}\\
		\hline
		\textbf{Injector} & \textbf{\#shot} & \textbf{$P_{av}$} & \textbf{$P_l$} & \textbf{$P_{cx}$} &\textbf{$P_h$}\\ 
		\hline\hline
		\multirow{2}{*}{NBI2} & 53577 & 150 & 12 & 105 & 33\\
		& 53605 & 111 & 11 & 73 & 27\\
		\hline
		\multirow{2}{*}{NBI1} & 54097 & 80 & 7 & 57 & 16\\
		& 24000 & 195 & 28 & 51 & 116\\ 
		\hline
	\end{tabular}
	\caption{Distribution of available power ($P_{av}$) between different channels: $P_l$ and $P_{cx}$ are the orbit and CX lost power respectively while $P_h$ is the power transferred to the plasma.}
	\label{tab:powerdis}
\end{table}

The balance between the available power ($P_{av}$) and the power distributed between the different channels is presented in table \ref{tab:powerdis}. Roughly, orbit losses ($P_l$) represent 10\% of the available power in all cases, while 70\% is lost by CX ($P_{cx}$), leaving a 20\% for heating ($P_h$). The exception is shot \#24000, for which only 26\% of the available power is lost by CX and 59\% is transferred to the thermal plasma. The reason for this is understood by examining the different neoclassical (orbits and collisions) and CX timescales, characterized by $\tau_{neo}$ and $\tau_{cx}$ respectively. As defined here, $\tau_{neo}$ is a typical lifetime of fast ions, bearing in mind that they can escape the plasma (orbits losses) or thermalize (slowing down). According to the simulations results, $\tau_{neo}\sim25$ ms is found for the NBI+ECRH plasmas and $\tau_{neo}\sim10$ ms for the NBI one, while $\tau_{cx}=[n_0\sigma_{cx}(v)v]^{-1}\sim1$ ms is obtained for a neutral density $n_{0}=10^{16}$ m$^{-3}$ and a fast-ion energy (E$_f$) of 15 keV. At this energy the CX cross section is $\sigma_{cx}=10^{-19}$ m$^2$ (see \cite{Riviere1971}). The smaller difference between timescales found in the NBI shot make up for the different behaviour in the mentioned power balance.

\subsection{NBCD calculation}
\label{sec:nbcd}

The beam-driven current is the combination of two different contributions: the current carried by the fast-ion population and the electron return current created by the response of the bulk electrons to the beam ions. The fast-ion current density can be calculated as
\begin{equation}
\label{eq:beamcurr}
J_{b\|}=e Z_{b}\int v_{\|}f(\mathbf{r},\mathbf{v})d\mathbf{v},
\end{equation}
where $e Z_{b}$ is the fast-ion charge, and $f(\mathbf{r},\mathbf{v})$ is the distribution function of fast-ions calculated with ASCOT. Following the results derived in \cite{Mulas_2022}, the total current driven by the beam can be written as
\begin{equation}
\label{eq:beamcurrfluxav}
J_{\|}^{nb}=J_{b \|}(1-A).
\end{equation}
Here, the term $1-A$ is the correction factor that modifies the fast-ion current due to the electron return current. It has been calculated by solving the Drift Kinetic Equation (DKE) for the electron distribution function modified by the presence of the fast-ion beam in the low-collisionality regime. The explicit expression of $A(s,\theta,\phi)$ is derived in \cite{Mulas_2022}. It is worth noticing that A depends on $Z_{eff}$ and on the precise magnetic geometry but not on the fast ion current itself since all dependence on $J_{b\parallel}$ is factored.

\begin{figure}[t]
\begin{center}
\includegraphics[width=0.35\textwidth]{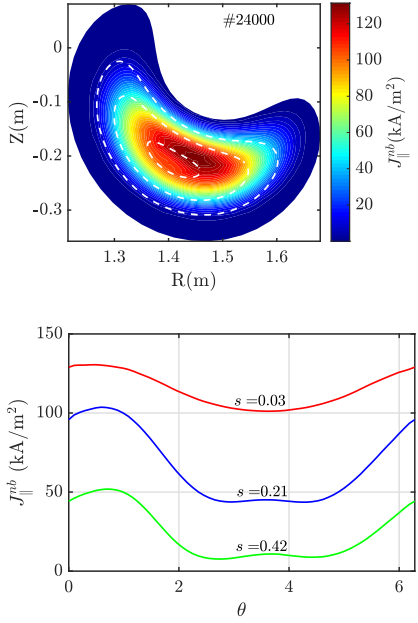}
\end{center}
\caption{On top, we show a 2D map of the current density driven by the NBI1 beam in shot \#24000. The poloidal current variations along the flux surfaces plotted in the 2D map (dashed white lines) are shown in the bottom panel.}
\label{fig:24000_jmap}
\end{figure}
\begin{figure}[h]
\begin{center}
\includegraphics[width=0.35\textwidth]{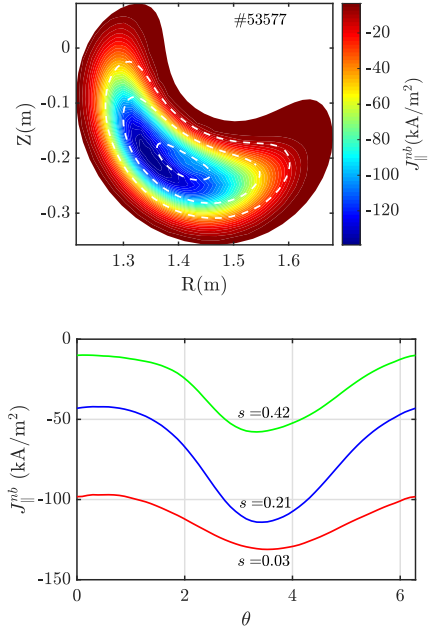}
\end{center}
\caption{Same as figure \ref{fig:24000_jmap} for NBI2 in shot \#53577. As expected for counter-injection, parallel current density is negative.}
\label{fig:53577_jmap}
\end{figure}

\begin{figure}[h]
\begin{center}
\includegraphics[width=0.45\textwidth]{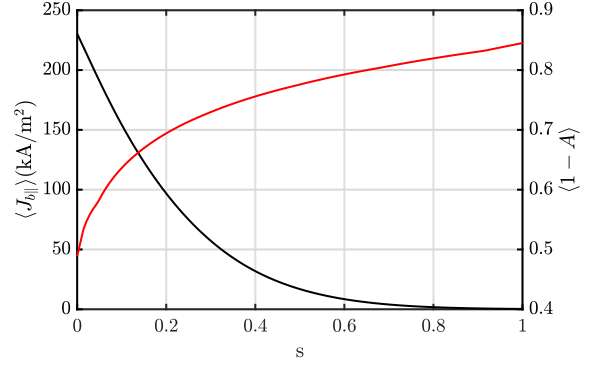}
\end{center}
\caption{Flux surface average of the fast ion current $J_{b\parallel}$ (black solid line) and the shielding factor $1-A$ (red solid line).}
\label{fig:jbF}
\end{figure}

The 2D maps of $J_{\|}^{nb}$ for shots \#24000 and \#53577, at a given toroidal angle, are shown in figures \ref{fig:24000_jmap} and \ref{fig:53577_jmap} respectively. As it was show in \cite{Mulas_2022}, neither the fast-ion density nor $J_{b||}$ are flux functions and the poloidal variation of the beam driven current density on selected surfaces is illustrated in the figures.  To show clearly how the shielding factor impacts the fast ion current, the flux surface average of both quantities, calculated for the case presented in figure \ref{fig:24000_jmap}, are plotted in figure \ref{fig:jbF}. 

The contribution of $J_{\|}^{nb}$ to the total toroidal current inside a flux surface is calculated as follows,
\begin{align}
\mathcal{I}_{nb}(s)&=\int{J_{\|}^{nb}\mathbf{b}\cdot d\textbf{S}_{\phi}}   \nonumber\\
&=\int_0^{s} ds^\prime\int_0^{2\pi}{J_{\|}^{nb}\frac{B^{\phi}}{|\mathbf{B}|}\sqrt{g}d\theta}  \nonumber\\
&=\int_0^{s} ds^\prime\frac{d\mathcal{I}_{nb}}{ds^\prime}
\end{align}

\begin{figure}[h]
	\begin{center}
		\subfigure[]{\includegraphics[width=0.45\textwidth]{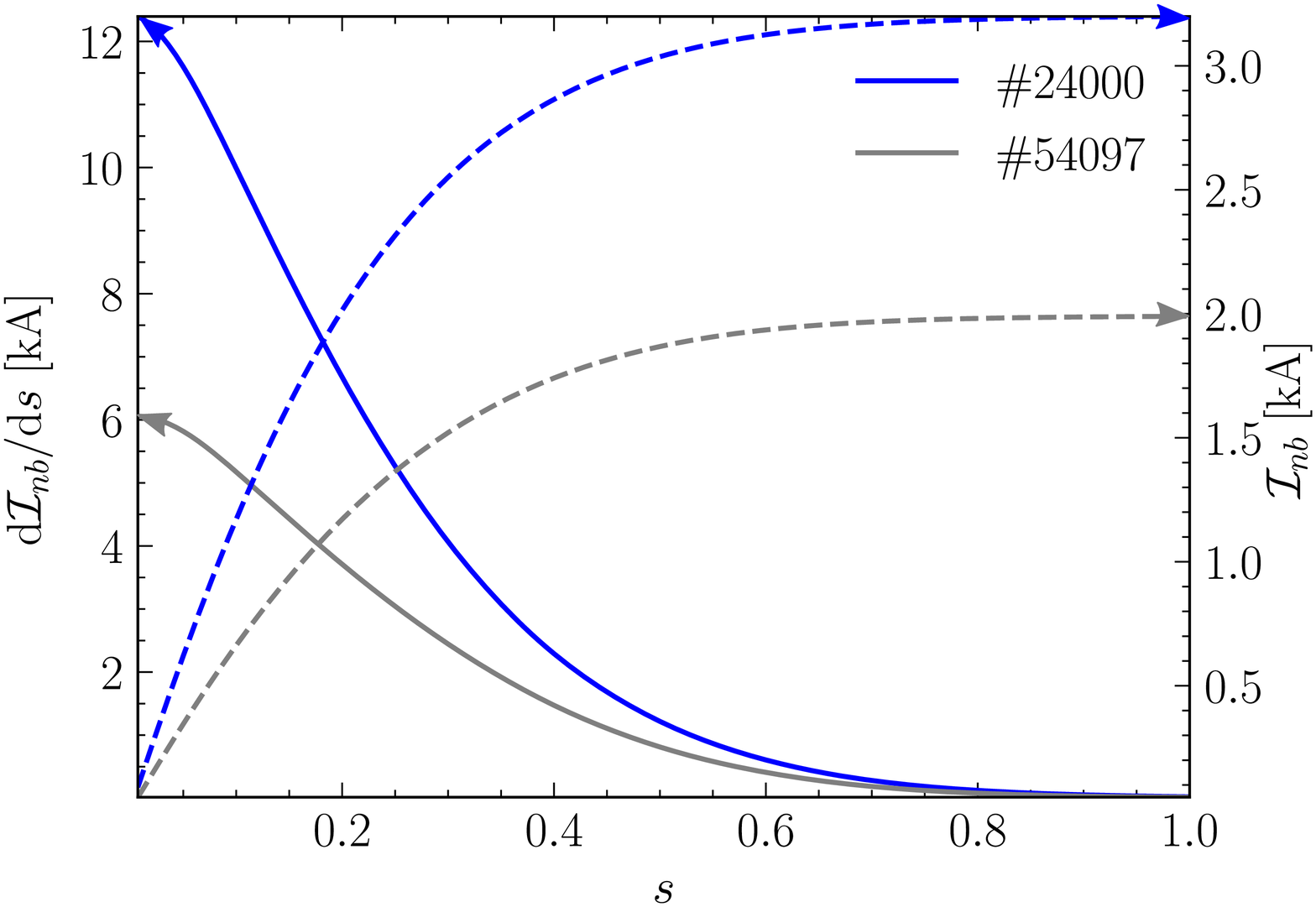}\label{fig:FIcurrent_NBI1}}
		\subfigure[]{\includegraphics[width=0.476\textwidth]{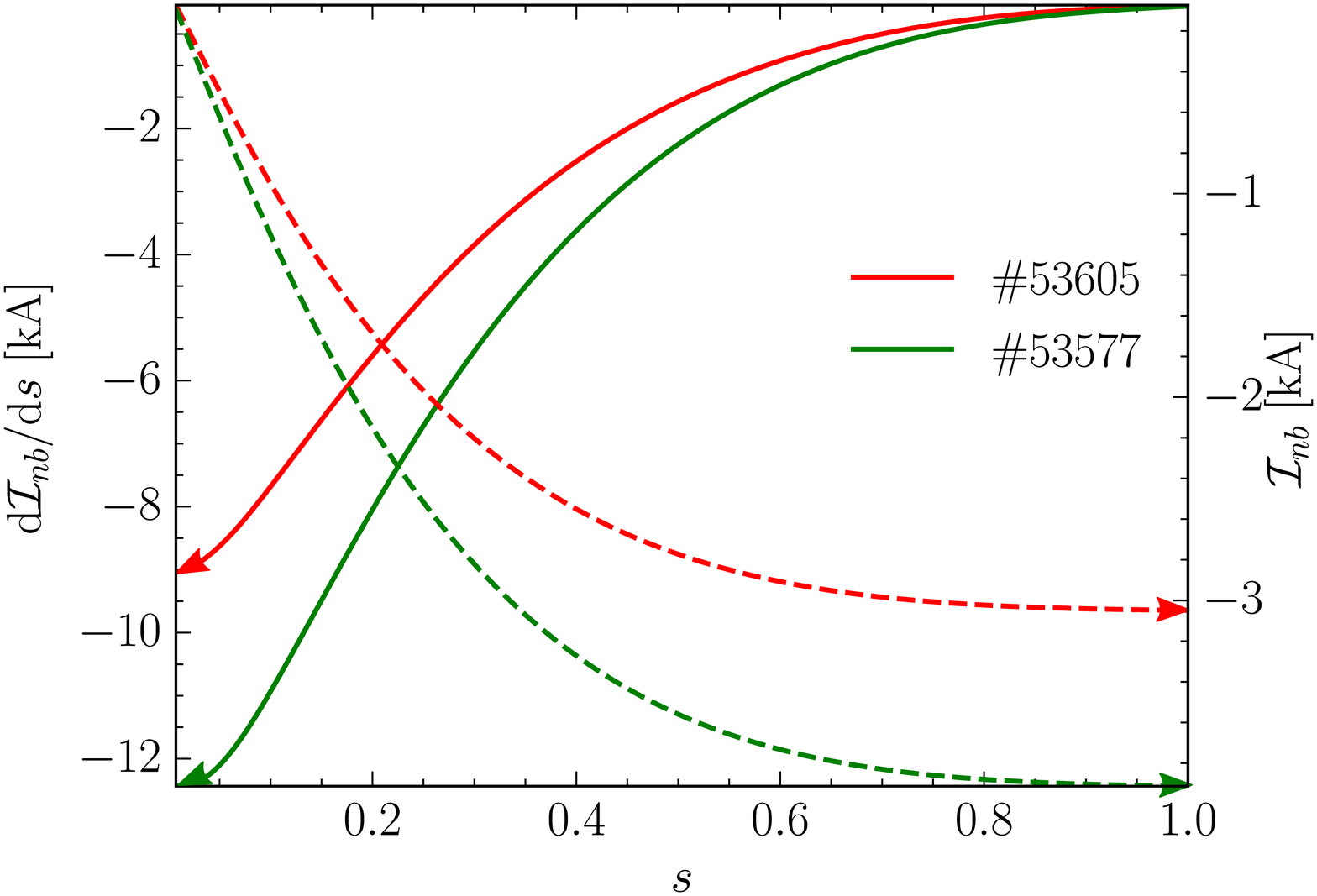}\label{fig:FIcurrent_NBI2}}\\        
	\end{center}
	\caption{NBCD density (solid lines) and integrated current (dashed lines) for the NBI1 \protect\subref{fig:FIcurrent_NBI1} and NBI2 shots \protect\subref{fig:FIcurrent_NBI2}.}
	\label{fig:NBIcurrents}
\end{figure}

\begin{figure}[h]
	\begin{center}
		\subfigure[]{\includegraphics[width=0.45\textwidth]{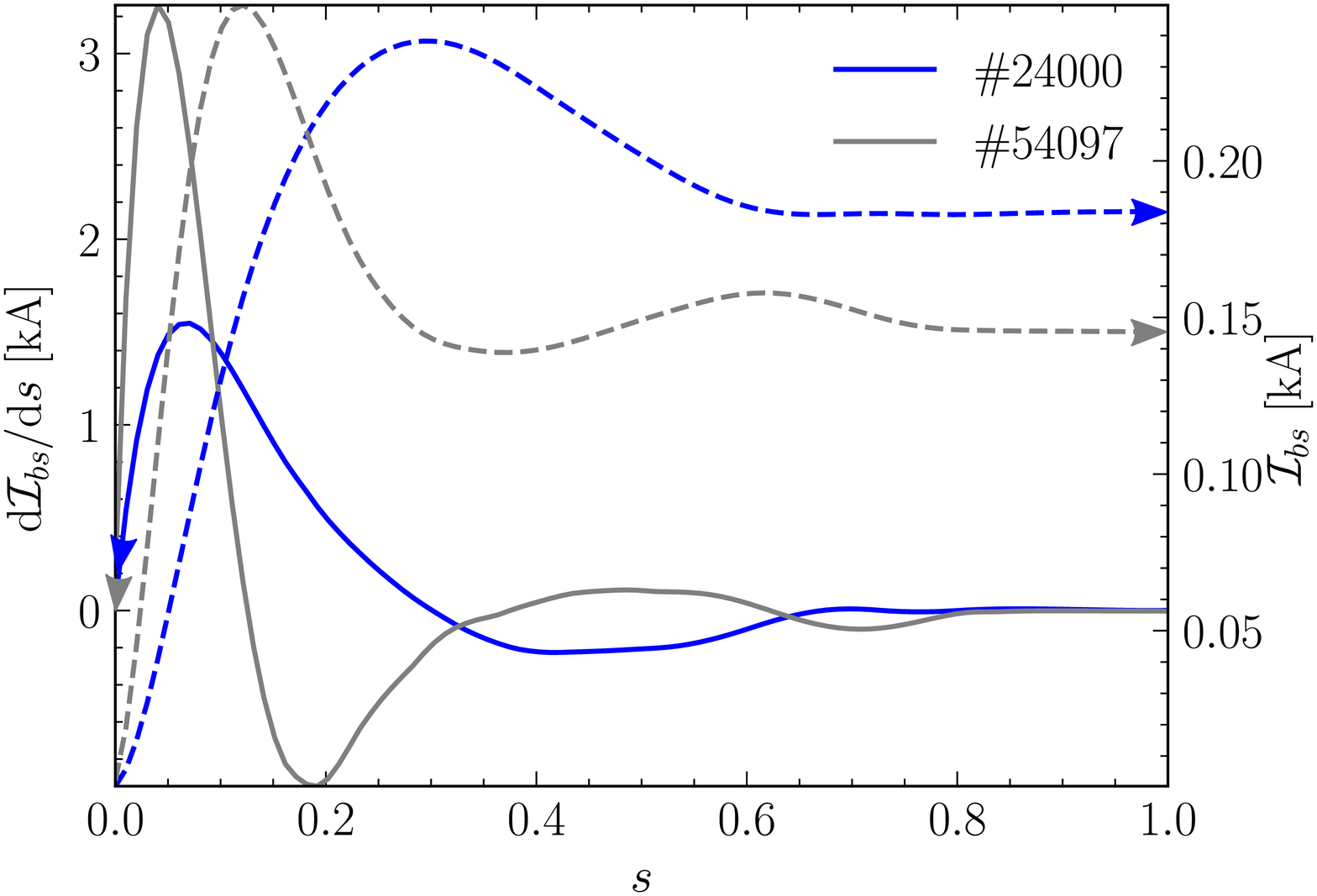}\label{fig:BScurrent_NBI1}}
		\subfigure[]{\includegraphics[width=0.46\textwidth]{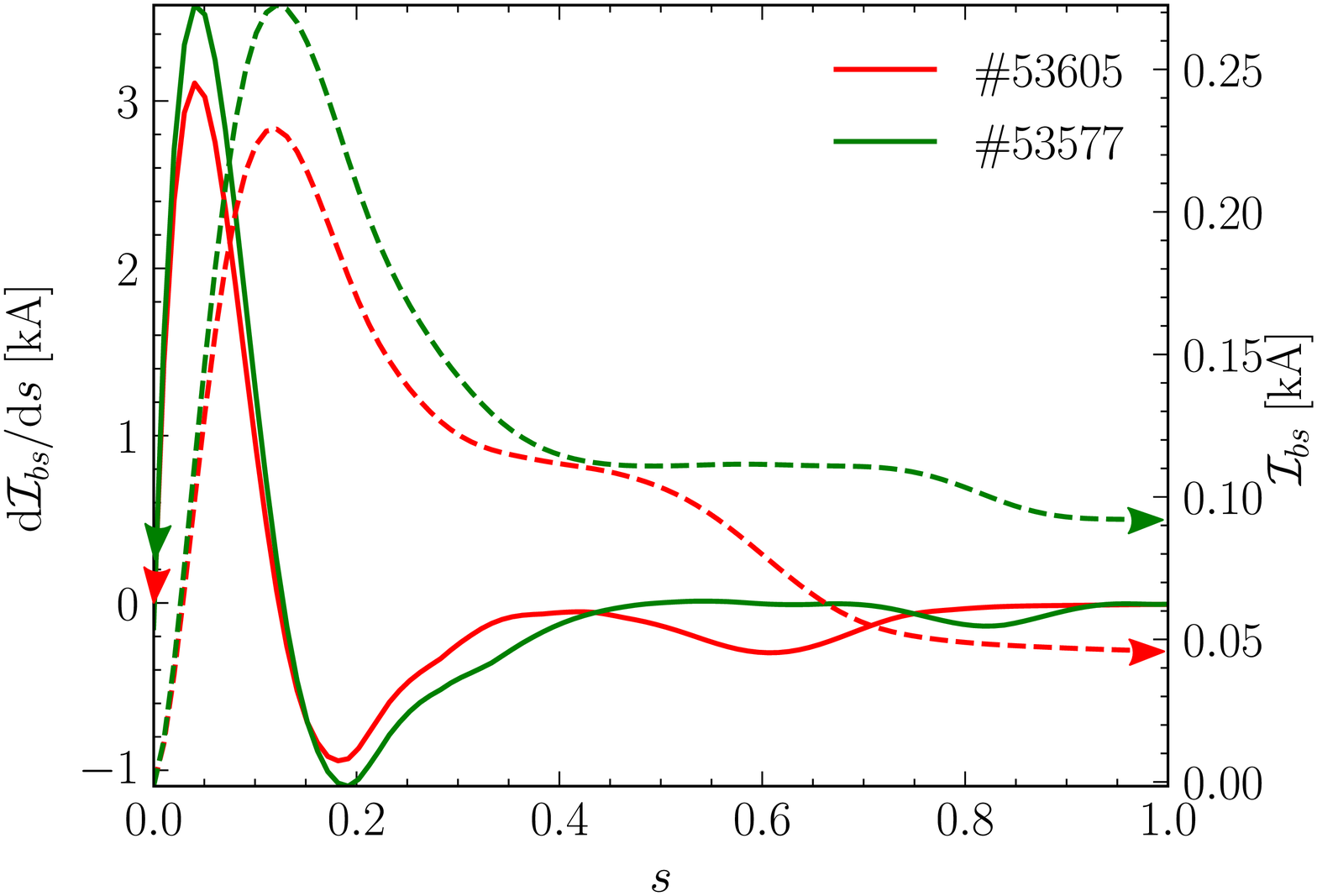}\label{fig:BScurrent_NBI2}}\\        
	\end{center}
	\caption{Bootstrap current density (solid lines) and cumulative current (dashed lines) for NBI1 shots \protect\subref{fig:BScurrent_NBI1}, and NBI2 shots \protect\subref{fig:BScurrent_NBI2}.}
	\label{fig:BScurrents}
\end{figure}

The neutral beam toroidal current density profiles ($d\mathcal{I}_{nb}/ds$) and the corresponding integrated currents inside a given flux surface ($\mathcal{I}_{nb}(s))$, are shown in figure~\ref{fig:NBIcurrents}. The total integrated currents produced by NBI injection in each of the studied cases, $I_{nb}\equiv\mathcal{I}_{nb}(s=1)$, are shown in table \ref{tab:currents}. As we stated in the introduction, we need to consider the contribution of the bootstrap current to the total toroidal plasma current. The radial profiles of bootstrap current density ($d\mathcal{I}_{bs}/ds$) and the corresponding integrated currents ($\mathcal{I}_{bs}(s)$), calculated with the code DKES, are presented in figure \ref{fig:BScurrents}. As for the case of $I_{nb}$, $I_{bs}\equiv\mathcal{I}_{bs}(s=1)$. The uncertainties of the theoretical $I_{nb}$ and $I_{bs}$ values have been calculated taking the different plasma profiles allowed within the error bars of the Thomson scattering diagnostic and, for the case of $I_{nb}$, considering an additional $\pm10$\% uncertainty in the neutral density profiles. These values of $I_{bs}$ also appear in table \ref{tab:currents}. The positive values of $I_{bs}$ oppose the $I_{nb}$ ones for the cases where NBI2 is used (negative NBCD), thus reducing the total amount of current in the plasma. On the contrary, $I_{bs}$ and $I_{nb}$ are both positive for the NBI1 cases. Anyway, in these particular cases, the contribution of bootstrap current to the toroidal plasma current is negligible.

\begin{table}[h]
\small
\centering
	\begin{tabular}{|c|c|c|c|c|}
    \cline{2-5}
    \multicolumn{1}{c|}{} & \multicolumn{4}{c|}{\textbf{Currents [kA]}}\\
    \cline{2-5}
    \multicolumn{1}{c|}{} & \multicolumn{2}{c|}{\textbf{NBI2}} & \multicolumn{2}{c|}{\textbf{NBI1}}\\
    \cline{2-5}
    \multicolumn{1}{c|}{}    & \textbf{\#53577} &  \textbf{\#53605} & \textbf{\#54097} & \textbf{\#24000}\\
    \hline
    \textbf{$I_{nb}$} & $-3.8\pm0.4$ & $-3.0\pm0.3$ & $2.0\pm0.2$ & $3.1\pm0.4$\\
    \hline
    \textbf{$I_{bs}$} & $0.1\pm0.1$ & $0.1\pm0.1$ & $0.2\pm0.1$ & $0.2\pm0.2$\\
    \hline
    \hline
    \textbf{$I_{th}$} & $-3.7\pm0.4$ & $-2.9\pm0.3$ & $2.2\pm0.2$ & $3.3\pm0.4$\\
    \hline
    \textbf{$I_{ni}$} & $-3.5\pm0.4$ & $-2.0$ & $0.6\pm0.1$ & $1.6$\\
    \hline
 \end{tabular}%
\caption{Computed NBCD (I$_{nb}$), bootstrap (I$_{bs}$), and total (I$_{th}$=I$_{nb}$+I$_{bs}$) calculated currents. The experimental asymptotic current (I$_{ni}$) is shown for comparison.}
\label{tab:currents}
\end{table}

Finally, following \ref{ap:a}, we obtain $I_{ni}$, the asymptotic value of the non-inductive toroidal current as measured by the Rogoswki coil. Whenever possible, an error bar for $I_{ni}$ has been given by averaging over highly reproducible shots obtained in the same experimental session. As it is shown in table \ref{tab:currents}, the calculated total toroidal current ($I_{th}=I_{bs}+I_{nb}$) shows a reasonable agreement for the case of the counter NBI injector (NBI2) while the experimental values are well below the calculated values in the case of the co-injector (NBI1). 

\section{Discussion}

The reason for the discrepancy between the experimental behaviour of NBI1 and NBI2 is not clear at the time of writing this paper. Differences in beam geometry can be ruled out in account of the internal calorimetric measurements. Beams of similar power on the V-calorimeter deposit similar power on the duct scrapers. Reionization losses are also similar in both injectors, as obtained by IR measurements \cite{Liniers_2013}. Nevertheless, shots with similar plasma parameters and similar available power, \#54097 for NBI1 and \#53605 for NBI2, exhibit quite different absolute values of plasma current, 0.6 kA and 2.0 kA respectively. We know from previous studies \cite{Mulas_2022} that prompt losses are about \%10 higher in the case of NBI1 but this can hardly justify the difference between observed currents. On the other hand, the results of the simulations are in rather good agreement with the observed plasma current driven by NBI2. Therefore, it is not unreasonable to think that NBI1 is driving less current than expected. 

A well-known effect that can have a noticeable impact on the confinement of fast ions is related to the excitation of Alfvén Eigenmodes (AEs) by the fast ions themselves. Fast ion transport may be enhanced when interacting with such types of instabilities. 

\begin{figure}[h]
	\begin{center}
		\subfigure[]{\includegraphics[width=0.455\textwidth]{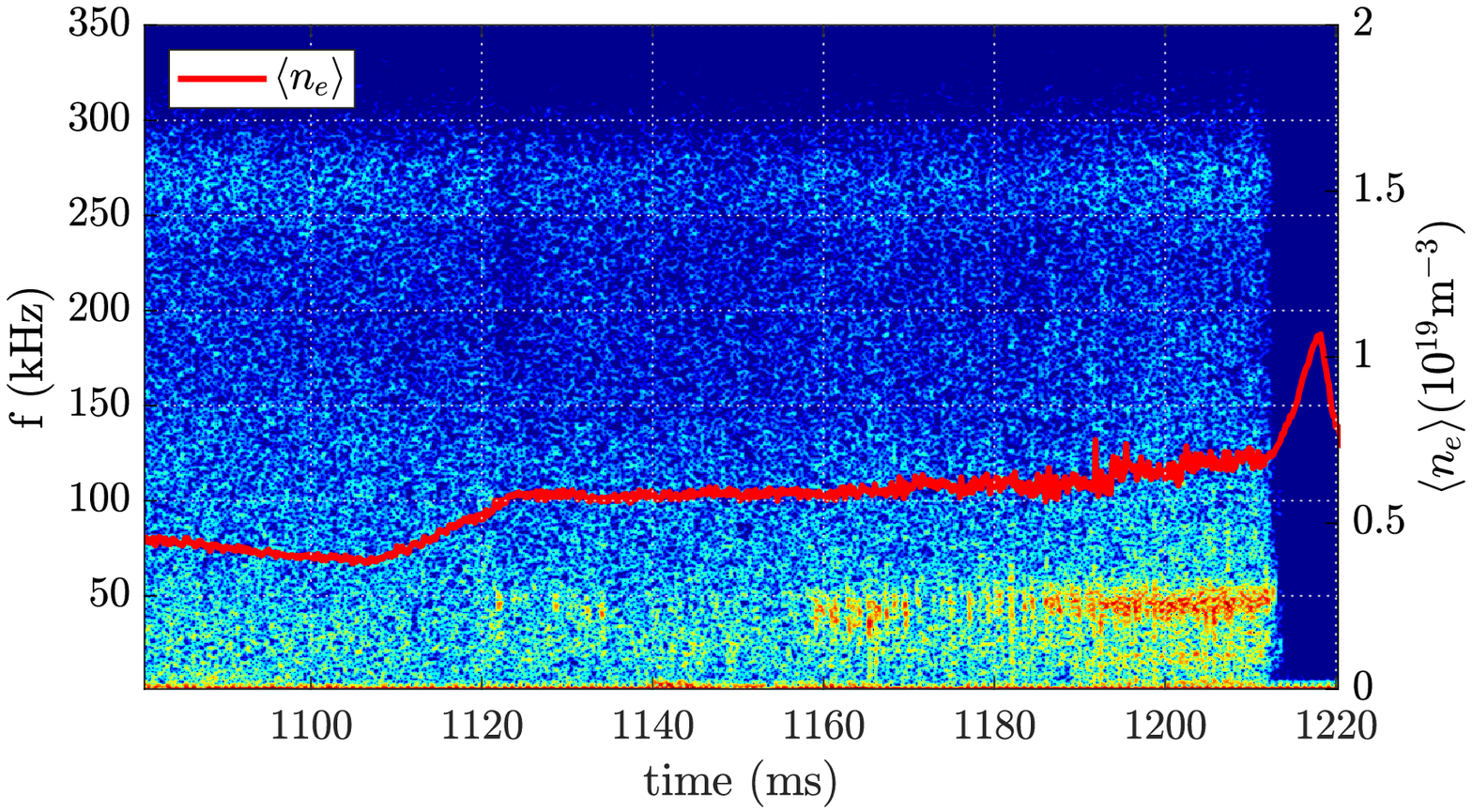}\label{fig:modes_NBI1}}
		\subfigure[]{\includegraphics[width=0.455\textwidth]{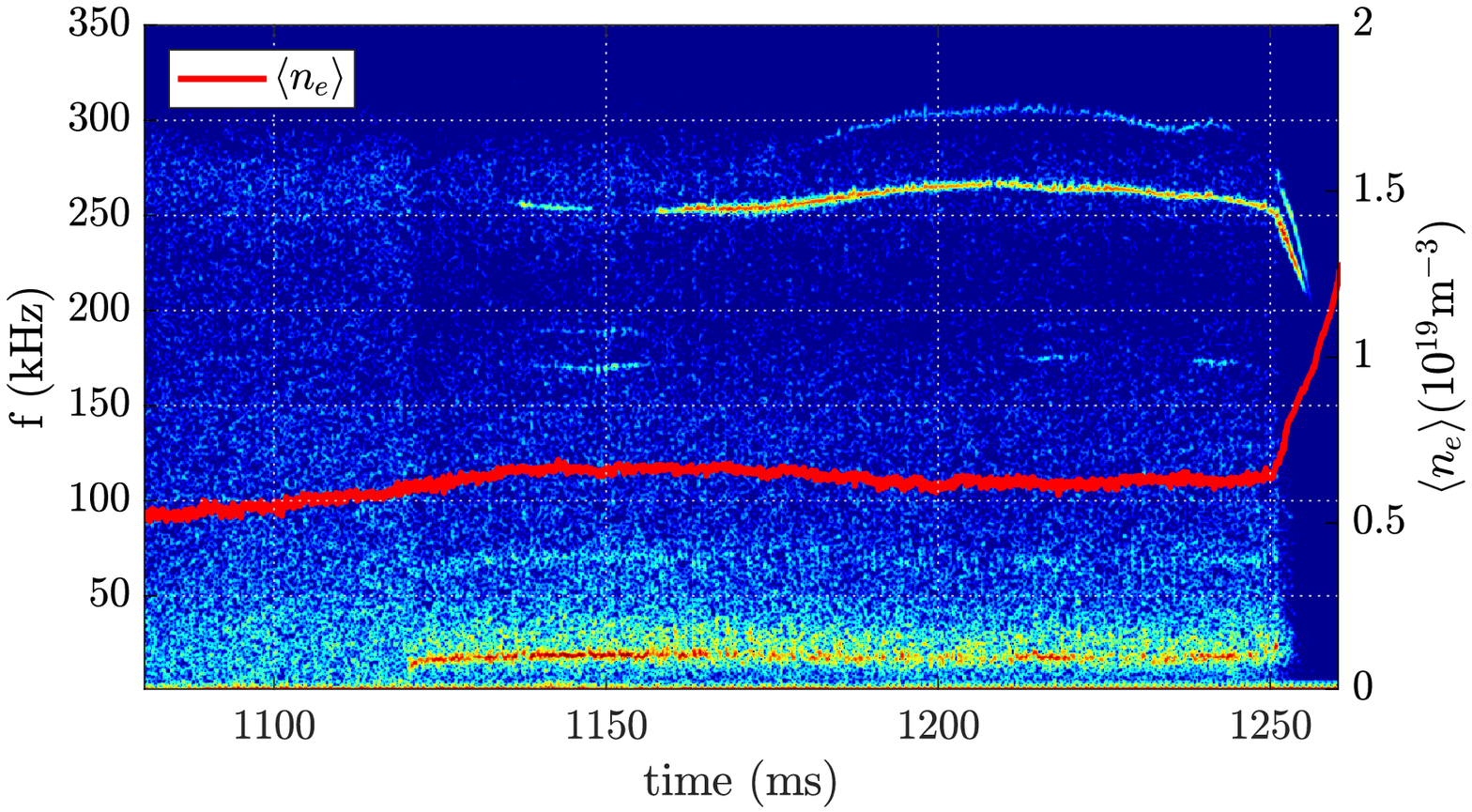}\label{fig:modes_NBI2}}\\     
	\end{center}
	\caption{Magnetic fluctuations spectrogram observed with NBI1 \protect\subref{fig:modes_NBI1} in shot \#54097, and NBI2 in shot \#53605 \protect\subref{fig:modes_NBI2}. Line density is over-plotted (red solid line).}
	\label{fig:aes}
\end{figure}

This effect is particularly important in tokamaks and is currently one of the main fast-ion research topics in such devices \cite {Ishikawa_2006, Garcia-Munoz_2011, Heidbrink_2014, Kramer_2017}. It could be speculated that this effect is playing a role here by lowering the efficiency of NBI1. However, NBI2 plasmas and not those of NBI1 are the ones that generally exhibit a much higher activity of Alfvén Eigenmodes. For instance, in the medium NBI power case, NBI1 is no longer able to excite AEs while NBI2 still does (see figure \ref{fig:aes}). We may take this as an indication that a lower pressure of fast ion is achieved with NBI1 without involving any effect related to AEs, thus supporting the previous argument; less fast-ion pressure is consistent with a lower current. However, the stronger AEs activity seen in NBI2 plasmas could also be explained by recalling that very different iota profiles are obtained with negative (NBI2) and positive (NBI1) current. Changes in the current may have a strong impact on the spectrum of shear Alfvén waves \cite{Cappa_2021}. Therefore, stronger AEs activity cannot be directly linked to higher fast ion pressure and simulation of AEs excitation are needed to clarify the observed behaviour. Actually, the best way to resolve the uncertainty is to measure the fast-ion losses using a fast-ion loss detector (FILD). Routine operation of the FILD detector will become possible in the next experimental campaign of TJ-II and the comparison of fast ion losses between NBI1 and NBI2 plasma will be the subject of future work. 

The amount of neutral hydrogen in the plasma is also an important quantity subject to uncertainty. However, the same procedure is used to estimate the population of neutral atoms in NBI1 and NBI2 plasmas and therefore these estimates should be valid unless other populations of neutral species, disregarded in the present simulations, were having an impact on the CX losses only when NBI1 is launched. A possible explanation is the different deposition around the device of the lithium used for wall conditioning. Very likely, the irregular distribution of lithium and the complex shape of the TJ-II vessel (see figure \ref{fig:apA2b} in \ref{ap:a}) can produce localized areas were lithium concentration in the wall is higher. In case the injection of NBI1 favors the desorption of lithium, due for instance to a higher lithium deposition in the duct of the injector, such an effect could occur. Charge exchange reactions involving neutral lithium are not included in ASCOT and therefore, even assuming that we had estimates of its 3D distribution, the effect that it may have on the slowing-down of fast ions can not still be quantified. 

Other possibilities for the observed behaviour need to be addressed. An hypothetical misalignment of the ECRH beam used for density control could produce unwanted ECCD, but no ECH is used in shot \#24000 and still less than half the predicted current is observed. Another possible source of error could be in the calculation of the bootstrap current, which showed disagreement with the measured values in \cite{Dinklage2018}. Furthermore, only its ion contribution has been validated in TJ-II ECRH plasmas \cite{Arevalo_2014}. However, the small amount of bootstrap current compared to the one driven by the beam implies that this would be only a minor correction. More sophisticated hypothesis, related to the impact on plasma transport produced by NBCD (through changes in the rotational transform) or by the beam perpendicular currents (presumably modifying the radial electric field) are now being considered but are currently out of the scope of the paper.

\section{Conclusions}

Despite the difficulties inherent in conducting NBCD experiments in TJ-II, i.e., reaching stable density NBI plasmas and achieving a reliable determination of the asymptotic toroidal current, the experimental results obtained have allowed the comparison with the predictions of the theoretical model. After reviewing the different possibilities that can account for the discrepancies observed when using the co-injector, we conclude that the model and methodology that we have followed are able to predict with good accuracy the shine-trough power and neutral beam current drive, provided CX reactions with lithium, influence of AEs activity or more exotic physics are not being at play. We also conclude that accounting for CX losses is essential in TJ-II plasmas and that disregarding this effect would produce much higher fast ion currents resulting in wrong NBCD predictions. Aiming at clarifying the origin of the discrepancy between the currents driven by both neutral beams, further experiments (field reversal for instance) with improved diagnostics capabilities are planned for the forthcoming campaign.        
\section*{Acknowledgements}

The authors are grateful to the members of the ASCOT code community for the helpful discussions and advice with the code. This work has been carried out within the framework of the EUROfusion Consortium, funded by the European Union via the Euratom Research and Training Programme (Grant Agreement No 101052200 — EUROfusion). Views and opinions expressed are however those of the author(s) only and do not necessarily reflect those of the European Union or the European Commission. Neither the European Union nor the European Commission can be held responsible for them.

\begin{appendix}
\section{Inductive and non inductive contributions to the toroidal plasma current}
\label{ap:a}

\begin{figure}[h]
\begin{center}
\subfigure[]{\includegraphics[width=0.45\textwidth]{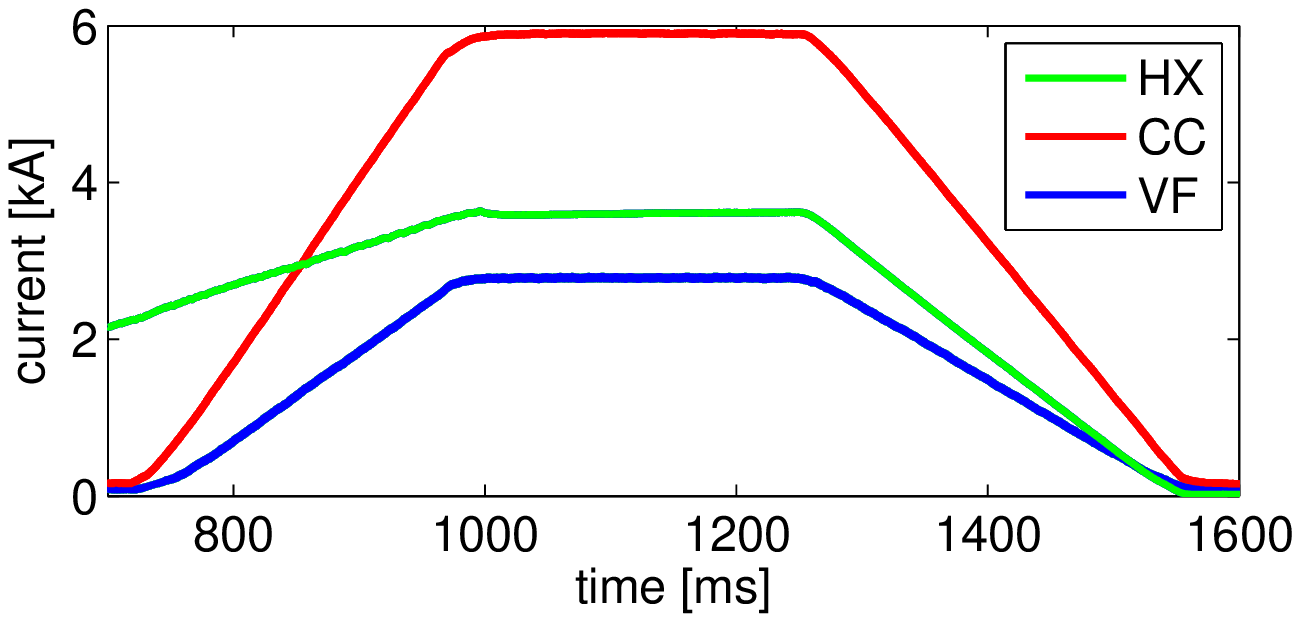}\label{fig:apA1a}}\\
\subfigure[]{\includegraphics[width=0.45\textwidth]{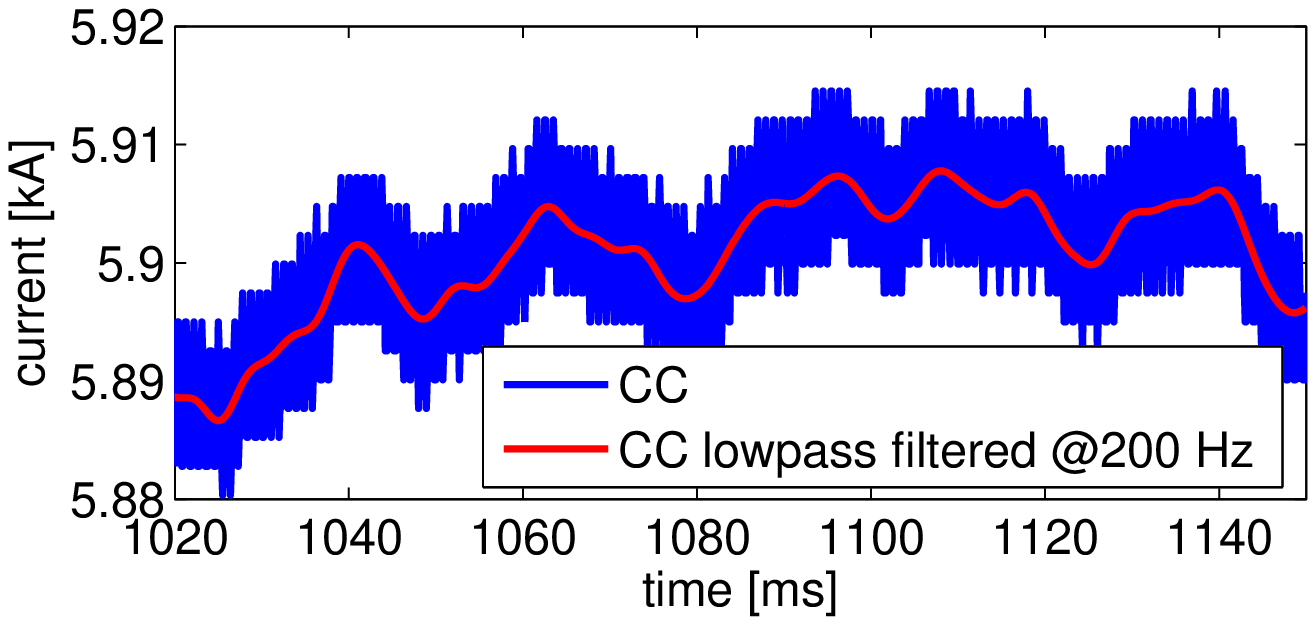}\label{fig:apA1b}}
\end{center}
\caption{Time traces of the different coil currents \subref{fig:apA1a} including current ramp-up and ramp-down. Enlarged view of the current evolution in the central coil \subref{fig:apA1b}. Applying a low pass filter helps separate high and low frequency contributions.}
\label{fig:apA1}
\end{figure}

In TJ-II, the time traces of the plasma current, measured by a Rogowski coil, exhibit large oscillatory deviations from the typical exponential behaviour expected from non-inductive sources (such as bootstrap, ECCD or NBCD) observed in other stellarator devices \cite{Klinger2019, Dinklage_2021, Watanabe_2002}. The currents of the vacuum-field coils present two superimposed oscillations at low and high frequency (see figure \ref{fig:apA1}). 

\begin{figure}[h]
\begin{center}
\subfigure[]{\includegraphics[width=0.45\textwidth]{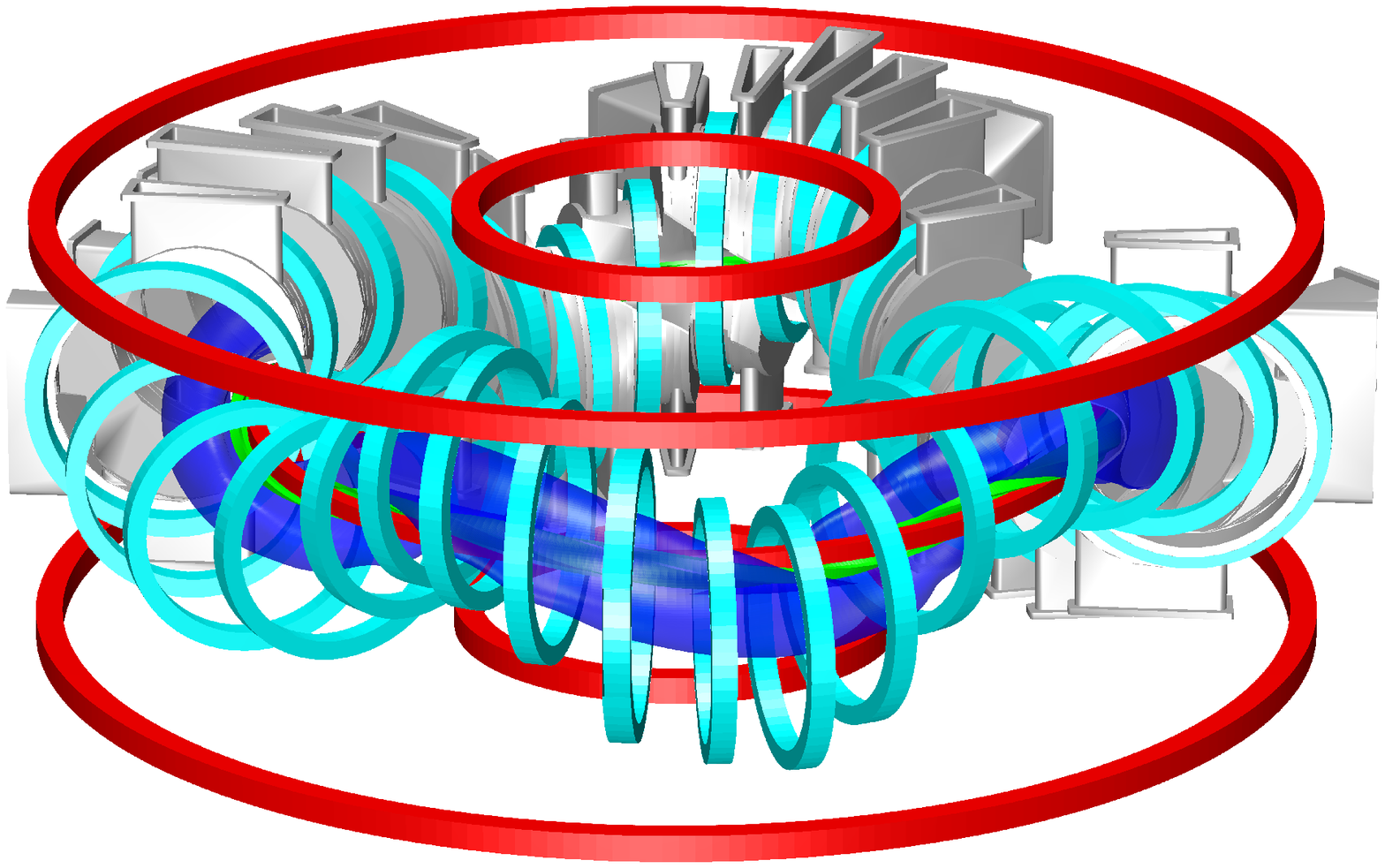}\label{fig:apA2a}}
\subfigure[]{\includegraphics[width=0.40\textwidth]{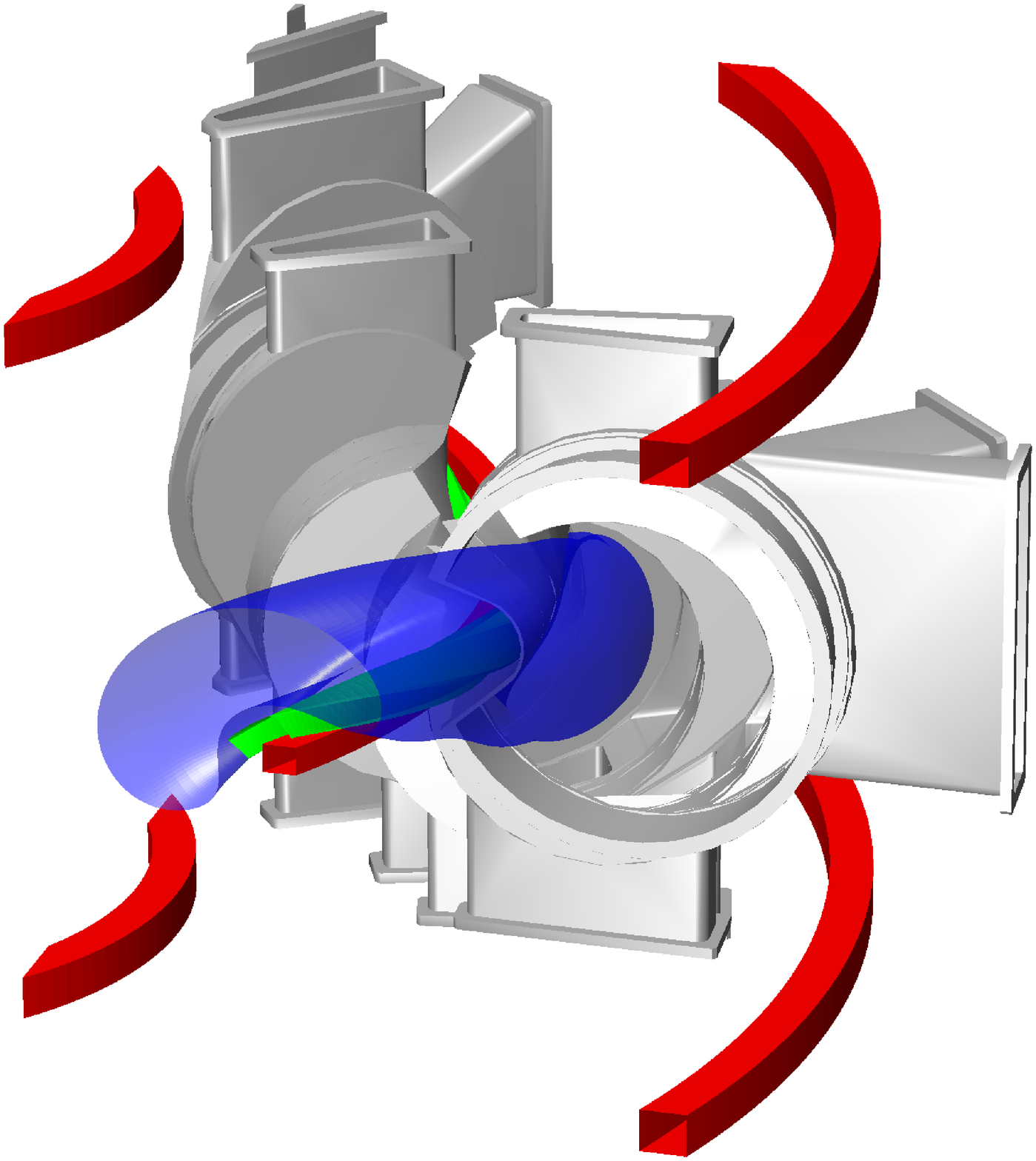}\label{fig:apA2b}}
\end{center}
\caption{General view of the TJ-II stellarator configuration coils \subref{fig:apA2a} and a detail view \subref{fig:apA2b} showing both the plasma in blue and the different coils: the central coil (CC), in red; the helical coil (HX) twisted around CC, in green; and the outer vertical field coils (VF), also in red.}
\label{fig:apA2}
\end{figure}

These oscillations and the proximity of the coils to the plasma (see figure \ref{fig:apA2}) produce a non-negligible inductive current in the latter. Assuming that the plasma and the coils can be described by a circuital model, the time evolution of the measured plasma current ($I_p$) is governed by
\begin{equation}
\tau_{LR}\frac{dI_p}{dt}+I_p =\sum_i\mu_i\frac{dI_c^i}{dt}+I_{ni}
\label{eq:dif}
\end{equation}
where $I_c^i$ are the currents in the different coils and $I_{ni}$ is the plasma current due to non-inductive sources. 
Being $R$ the plasma resistance and $L$ the plasma self-inductance, the time constant is defined as $\tau_{LR}\coloneqq L/R$ and the induction coefficients as $\mu_i\coloneqq M_i/R$, where $M_i$ are the inductive coupling coefficient between the plasma and the vacuum-field coils. Although the inductive oscillations of $I_p(t)$ could be calculated using the coefficients $\mu_i$, the unknown finite-volume distribution of $I_p$ inside the plasma prevent us from obtaining a reliable estimation. Instead, a minimization procedure has been applied to find the appropriate values of the coefficients so that equation \ref{eq:dif} is fulfilled.

A simple solution of equation \ref{eq:dif} exists when $\tau_{LR}$ and $I_{ni}$ are approximately constant, and either the ripple of the coil currents can be neglected ($dI_c^i/dt\approx0$) or the inductive couplings vanish ($M_i\approx 0$). In this case, the well-known exponential evolution is recovered,
\begin{equation}
I_p^c(t)=(I_p(0)-I_{ni})e^{-t/\tau_{LR}}+I_{ni}
\label{eq:exp}
\end{equation}
and the plasma current asymptotically approaches $I_{ni}$. When both coupling coefficients and $dI_c^i/dt$ cannot be neglected, as it is the actual case, it is helpful to write eq. \ref{eq:dif} in its integral form and then define the time dependent residual of eq. \ref{eq:dif}, $\mathcal{R}(t)$, as
\begin{align}\label{eq:int}
\mathcal{R}(t)\coloneqq I_p(t)&+\frac{1}{\tau_{LR}}\int_0^tI_p(t')dt'\\ \nonumber
   &-\sum_i\eta_iI_c^i(t)-\frac{I_{ni}t}{\tau_{LR}}-\varphi
\end{align}
where $\eta_i\coloneqq\mu_i/\tau_{LR}$ and $\varphi\coloneqq I_p(0)-\Sigma_i \eta_i I_c^i(0)$ is an integration constant. In \ref{eq:int}, $I_{ni}$ is assumed constant in the time interval under study. 

\begin{figure}[h]
\begin{center}
\includegraphics[width=0.48\textwidth]{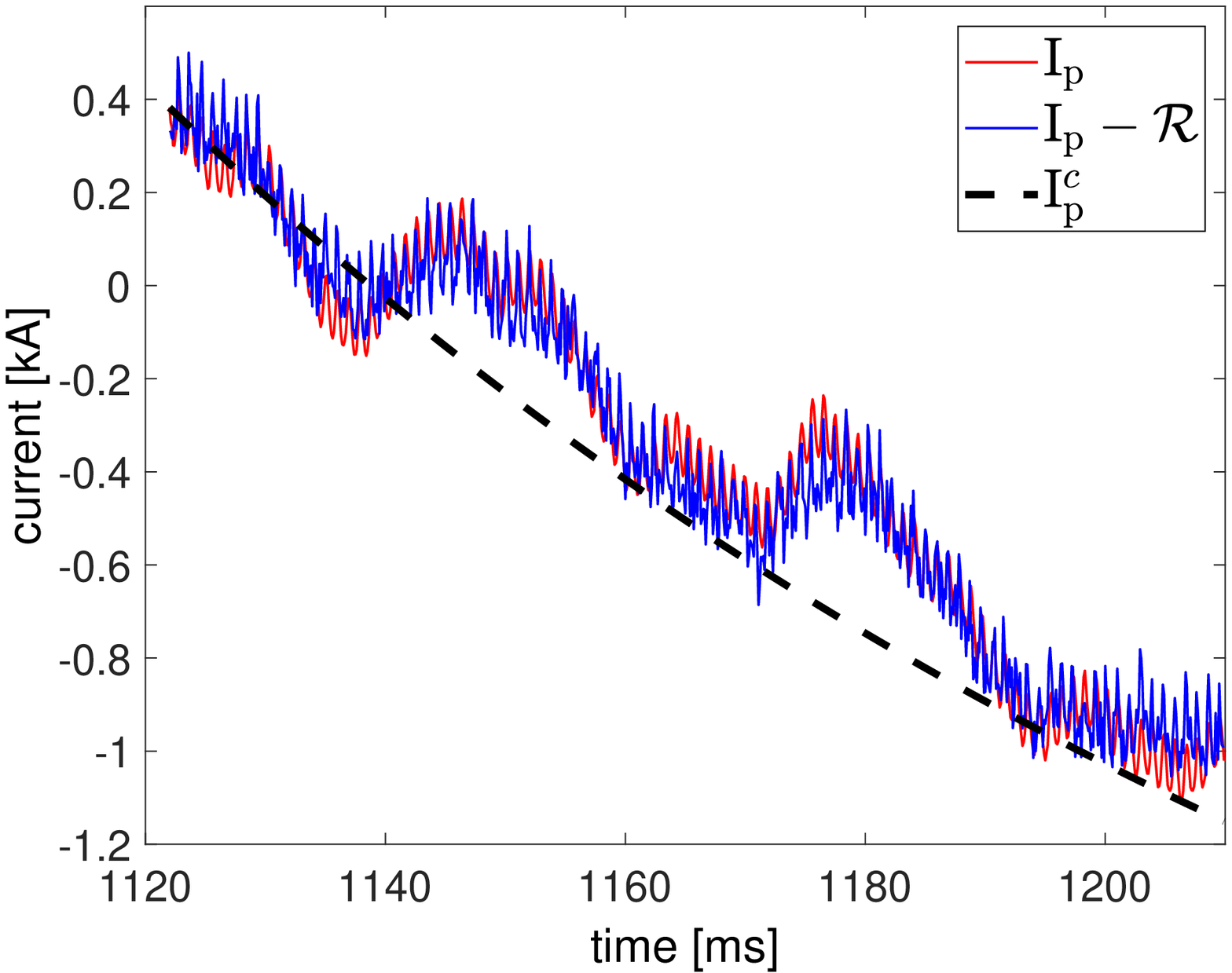}
\end{center}
\caption{Measured current $I_p$ (red line) and $I_p-\mathcal{R}$ (blue). The inferred exponential behaviour ($I_p^c$) is also depicted (black dashed line).}
\label{fig:optres}
\end{figure}

Since $I_p(t)$ and $I_c^i(t)$ are known, an optimization can be performed on the rest of parameters, namely $\tau_{LR}$, $\eta_i$, $I_{ni}$ and $\varphi$. The functional to be minimized is then $\sum_k|\mathcal{R}(t_k)|$. This also leads to the determination of $R$, $L$ and $M_i$, provided one of them is known. Figure \ref{fig:optres} shows the result of this process applied to an NBI discharge in a time window in which both the density and temperature were constant enough to ensure also a nearly constant $\tau_{LR}$ and $I_{ni}$. The red line plots the toroidal current measured by the Rogowski coil while the blue line depicts the rest of the terms in $\mathcal{R}$ evaluated with the parameters given by the optimization. Both time traces should be equal at all times in order for equation \ref{eq:dif} to be fulfilled. As shown in the figure, both traces show a very similar behaviour. Since $I_{ni}$ and $\tau_{LR}$ are given by the optimization, they can be used to estimate, according to \ref{eq:exp}, the time evolution of the plasma current as if there was no inductive contribution. This is also represented in Figure \ref{fig:optres}. 
\end{appendix}

\section*{References}
\bibliographystyle{unsrt}
\bibliography{mibiblio}

\end{document}